\documentclass[a4paper,twocolumn,superscriptaddress,11pt,accepted =2019-11-25]{quantumarticle}
\pdfoutput=1
\usepackage[utf8]{inputenc}
\usepackage[english]{babel}
\usepackage[T1]{fontenc}
\usepackage{amsmath, graphicx,amssymb,enumerate,verbatim, relsize,multirow}
\usepackage{hyperref}
\usepackage{tikz}
\usepackage{lipsum}

\usepackage[normalem]{ulem}
\usepackage[numbers,sort&compress]{natbib}

\renewcommand{\theenumi}{\Alph{enumi}}
\usepackage[inline]{enumitem}

\newtheorem{remark}{Remark}

\newcommand{\beq}{\begin{equation}}
\newcommand{\eeq}{\end{equation}}
\newcommand{\barr}{\begin{eqnarray}}
\newcommand{\earr}{\end{eqnarray}}
\newcommand{\tr}{\textrm{tr}}
\newcommand{\bra}[1]{\langle #1|}
\newcommand{\ket}[1]{|#1\rangle}

\newcommand{\proj}[1]{|#1\rangle\!\langle#1|}

\newcommand{\inner}[2]{\langle #1 | #2 \rangle} 
\newcommand{\ident}{\mathbb{I}}
\newcommand{\chan}[3]{\mathlarger{#1}^{\mathsmaller{#3}}_{\mathsmaller{#2}}}

\newtheorem{definition}{Definition}
 
\newtheorem{theorem}{Theorem}

\newcommand{\proof}{\textbf{Proof - }}
\newcommand{\proofend}{$\Box$}

\newcommand{\smallunderscore}{\textscale{.5}{\textunderscore}}

\newcommand{\oA}{A}
\newcommand{\oB}{B}

\usepackage{float}
\usepackage{soul, changepage}
\newcommand\T{\rule{0pt}{0.99cm}}       % Top strut
\newcommand\B{\rule{0pt}{0.65cm}} % Bottom strut

\newcommand{\PP}{\mathbf{p}}

\begin{document}

\title{Exploring the limits of no backwards in time signalling}

\author{Yelena Guryanova}
\affiliation{Institute for Quantum Optics and Quantum Information (IQOQI), Boltzmanngasse 3, Vienna 1090, Austria}
\author{Ralph Silva}
\affiliation{Group of Applied Physics, University of Geneva, Chemin de Pinchat 22,  CH-1211, Geneva 4, Switzerland}
\affiliation{Institute for Theoretical Physics, ETH Z\"urich, Wolfgang-Pauli-Str. 27, Z\"urich, Switzerland}
\author{Anthony J. Short}
\affiliation{H. H. Wills Physics Laboratory, University of Bristol, Tyndall Avenue, Bristol, BS8 1TL, United Kingdom}
\author{Paul Skrzypczyk}
\affiliation{H. H. Wills Physics Laboratory, University of Bristol, Tyndall Avenue, Bristol, BS8 1TL, United Kingdom}
\author{Nicolas Brunner}
\affiliation{Group of Applied Physics, University of Geneva, Chemin de Pinchat 22,  CH-1211 Geneva 4, Switzerland}
\author{\linebreak Sandu Popescu}
\affiliation{H. H. Wills Physics Laboratory, University of Bristol, Tyndall Avenue, Bristol, BS8 1TL, United Kingdom}

\begin{abstract}
We present an operational and model-independent framework to investigate the concept of no-backwards-in-time signalling. 
We define no-backwards-in-time signalling conditions, closely related to the spatial no-signalling conditions. These allow for theoretical possibilities in which the future affects the past, nevertheless without signalling backwards in time. This is analogous to non-local but no-signalling spatial correlations. Furthermore, our results shed new light on situations with indefinite causal structure and their connection to quantum theory.  
\end{abstract}

\maketitle

\section{Introduction}

One of the most routine observations that we make about our world is that we cannot signal backwards in time. So ubiquitous is this understanding that it is often taken as one of the basic laws of Nature\footnote{In principle, general relativity seems to allow for closed time-like curves, but there is currently no evidence of this in Nature.}. At first glance, this remark seems straightforward. However, as we show here, in probabilistic theories such as quantum mechanics, the consequences of such an assertion are far more involved. In fact, we will see that there is a surprising amount of liberty: some theories even allow the future to affect the past, nevertheless without signalling backwards in time.

Here, we consider the general properties of all theories which do not allow backwards-in-time signalling;  in some sense these properties are the temporal equivalent of the  no-signalling conditions that have played a  central role in the study of non-locality \cite{Bel64,pr1994,BellReview}. The fact that the future can affect the past without signalling backward in time is then akin to the phenomenon of non-locality, in which correlations can be non-local yet do not lead to superluminal signalling. 

In the case of non-locality, studying the most general correlations consistent with no-signalling, and the realisation that these were stronger than quantum correlations, led to a revolution in our understanding. On one hand such correlations might actually exist in nature, for example in some exotic contexts such as quantum gravity or in a future theory beyond quantum theory. This research is then a guide for what one should  look for.  On the other hand, if they do not exist, it is important to discover the  physical principles which rule them out. This led to the search for principles bounding quantum correlations \cite{popescureview} and the study of quantum theory `from the outside', renewed interest in generalised probabilistic theories \cite{barrett2007} and theories beyond quantum theory, and helped forge the device independent approach to quantum information science \cite{acin2007}. Here we wish to explore a similar idea for temporal correlations. 

Various other aspects of causality have also been studied recently, exploring the most general causal structures which might be possible in Nature \cite{Aharonov90,giulioIndefinite,Hardy07,Oreshkov2012,Leifer13,Baumeler14,Dominic16, DAriano2016, Perinotti2016}. Here we use our results on no-backwards-in-time-signalling to shed further light on some of these --- in particular on the process matrix formalism for indefinite causal structures \cite{Oreshkov2012,Branciard16a,Araujo1,Oreshkov16,Abbott16,Baumeler16}, and on pre- and post-selected quantum states \cite{ABL,Aharonov1991,Aharonov,multi,Ralph2014}, two approaches which were recently shown to be related \cite{Silva2017}. An interesting and surprising finding is that there are pre- and post-selected quantum states involving non-trivial post-selections that nevertheless do not lead to signalling backward in time and that the set of these states precisely correspond to process matrices.
  
\section{No backwards in time signalling}
Consider first a situation involving a single party, Alice, who performs the following procedure. At some point, a system enters her laboratory. She performs a measurement on it and obtains an outcome $a$. Alice then receives a classical random variable $x$ from outside the laboratory. She uses $x$ (and potentially also $a$)  to pick among a set of transformations that she then applies to the system. For example, Alice might apply some unitary transformation, or discard the system and prepare a completely new one, depending on the value of $x$. Finally, Alice sends the system out of her laboratory. We can describe the probability that Alice obtained the outcome $a$ given that she later received the input $x$ via the conditional probability distribution $p(a|x)$.

Now, if $p(a|x)$ depends on $x$, that would mean there is backwards in time signalling. More precisely, we assume that the input $x$ is a random variable, chosen by an external party who has no access to either the result $a$ or the input system (or anything else correlated to them). Under this assumption, the only way for there to be a dependence of $a$ on $x$ is through backwards-in-time signalling. 

Mathematically, we therefore define the \emph{no-backwards-in-time-signalling (NBTS) condition}\footnote{See also the note added at the end of the paper.} to be

 \beq p(a|x)=p(a).\label{causality_one}\eeq
which states that the probability of obtaining the outcome $a$ cannot depend on the input $x$ received later and which is used to choose which transformation is made  (this is similar to the causality condition defined in \cite{Chiribella2010}).

There are two important points that must be mentioned. First, note that if we would have allowed Alice to choose for herself the value of $x$, then it would be possible to have $p(a|x)\neq p(a|x')$ without any backward-in-time signalling. For example, consider the simple situation where Alice simply chooses $x = a$. Although the result $a$ is then correlated with $x$, there is clearly nothing backwards-in-time happening here. This demonstrates why it is essential to assume that $x$ is generated externally and independently of $a$ in order to be able to draw any conclusion about backward-in-time signalling. 

Secondly, as far as we can tell, there is no way to guarantee that the $x$  generated by the external party is really freely chosen, and has no `hidden' dependence on $a$ or the input system. One could take many precautions, to make the assumption as reasonable as possible but, as far as we can tell, it cannot be ruled out. However, it is not our objective here to discuss how to guarantee this.  Rather in the rest of the paper  we assume that this is the case and discuss its consequences. This is similar to the measurement-independence assumption in Bell's theorem, whereby the measurement settings must be independent of the hidden variables of the source, and is an implicit assumption that is generally accepted without further discussion.

Now, since any probability distribution satisfying \eqref{causality_one} can be achieved in either quantum theory or classical probability theory, we see that there is nothing particularly interesting in the above situation involving only a single party. However, as we shall see next, things becomes more subtle in situations involving two or more parties. 

Consider two parties, Alice and Bob, each of whom is in their own laboratory.  As before, at some point a system enters Alice's laboratory and she makes a measurement on it, with outcome $a$. She then receives a classical input $x$ from outside the laboratory. Next she performs some transformation on her system, which may depend upon $x$ and $a$, before sending the system out of her laboratory. Similarly, at some point a system enters Bob's laboratory. Again, he performs a measurement on it with outcome $b$, before receiving  a classical random variable $y$ from outside. He then performs a transformation on his system, which may depend on $y$ and $b$, before sending the system out of the lab. As above, we assume that the variables $x$ and $y$ are freely chosen by external parties who have no access to $a$ and $b$ or the input system(s). Depending on the relative timing of Alice and Bob's actions, it could be that the system that enters Alice's laboratory is the same as the one that left Bob's laboratory; in this case Bob could have affected its state and therefore the outcome $a$. Importantly however, we assume that Bob does not communicate to Alice in any other way during the experiment (or vice versa). 

\begin{figure}[t!]
\begin{center}
\includegraphics[width=1\columnwidth]{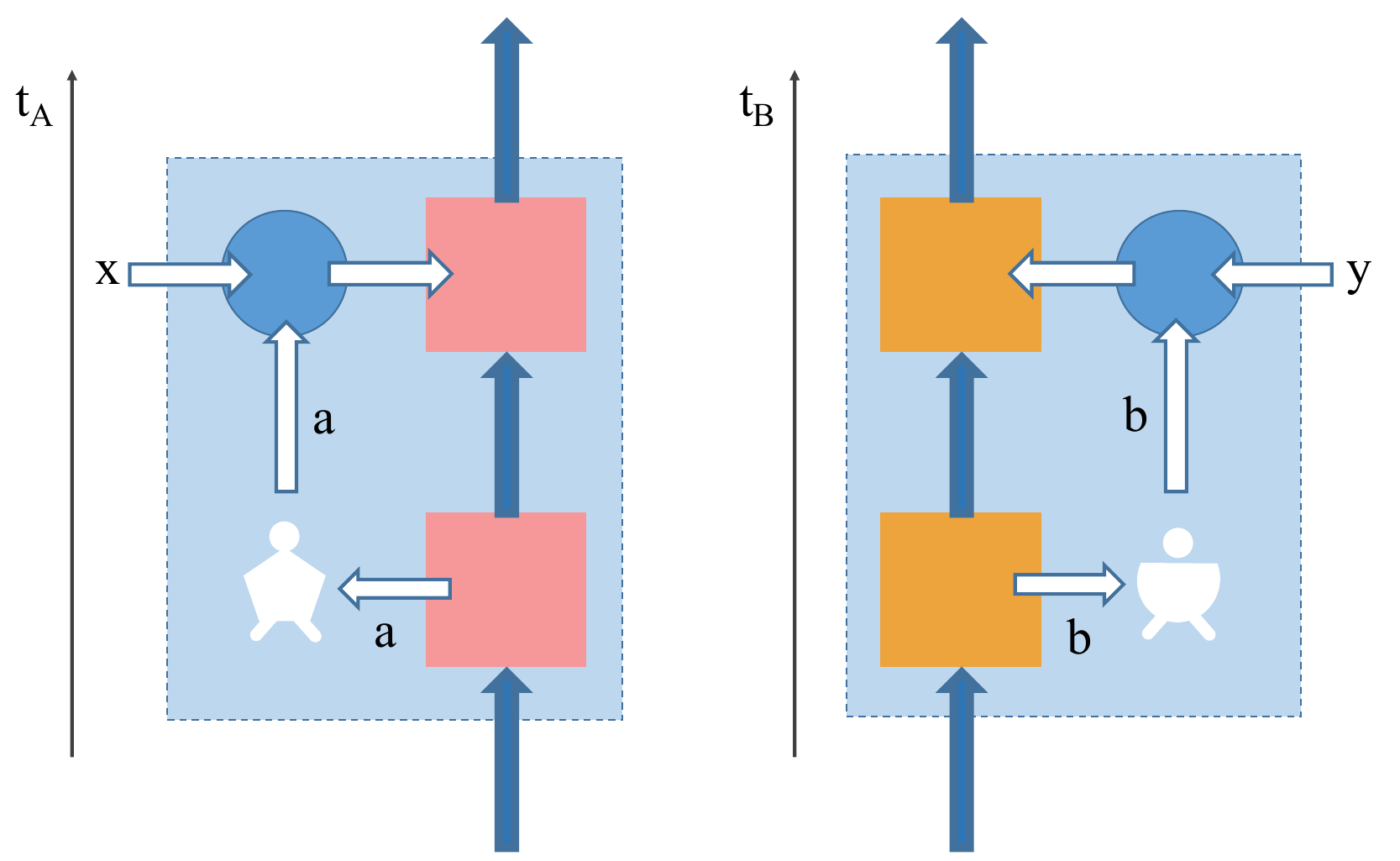}
\caption{An operational framework for investigating the no-backwards-in-time signalling conditions. Alice and Bob have separate closed laboratories. Each of them performs a measurement, and then receives a classical variable, generated freely outside the laboratory. They then perform a transformation on their system, depending on the externally generated classical variable and their measurement outcome, before sending the system out of the laboratory. Inside each laboratory time flows in the usual fashion. However, the assumptions about the relative time order of Alice's experiment versus Bob's vary from case to case as explained in the text.}
\label{fig:NBTsetup}
\end{center}
\end{figure}

The situation is then described by the joint conditional probabilities $p(a,b|x,y)$. See Fig.~\ref{fig:NBTsetup}.

Of course, as far as Alice is concerned, all she sees is the outcome $a$ (and the input $x$), and the probability for her to obtain an outcome is just determined by the marginal probability
\beq p_A(a|x,y):=\sum_b p(a,b|x,y).\eeq
Similarly, all Bob sees is $b$ and $y$, and Bob's outcomes will be distributed according to the marginal probability 
\beq p_B(b|x,y):=\sum_a p(a,b|x,y).\eeq

The question is how to write the NBTS conditions now. We distinguish various scenarios which differ in the assumptions about the {\it relative timing} between the experiments done by Alice and Bob.  All of them, however, have a main element in common: We can definitely assume time ordering inside each lab. That is, Alice performs her measurement and obtains the outcome $a$ before $x$ is delivered to her. Therefore, to avoid backwards-in-time-signalling we demand that Alice's outcome marginal probability is independent of $x$
\beq p_A(a|x,y)=p_A(a|y). \label{eq:NBTSA} \eeq
Similarly, for Bob we have
\beq p_B(b|x,y)=p_B(b|x).\label{eq:NBTSB}  \eeq

Coming now to the differences, there are two `extremal' cases, which reflect our knowledge about the relative timing between Alice and Bob's laboratories:

\begin{itemize}
\item \textbf{Definite relative timing}: 
In one case we have full knowledge about the relative timings between the laboratories.  
A particularly interesting situation is when Alice's and Bob's actions happen in parallel -- i.e. such that it is impossible that the same system passes through Alice's and Bob's labs, since there is not enough time for a system to travel from Alice's lab to Bob's lab, or vice versa
\footnote{Note that this possibility arises both in a relativistic setting, where the system cannot travel faster than the speed of light, or in a nonrelativistic setting (with infinite speed propagation of the system) as long as the two experiments overlap in time.}. 
Hence the probability of Alice's outcome cannot depend on either $x$ or $y$ and so
\beq p_A(a|x,y)=p_A(a). \label{eq:NBTSA2} \eeq
Similarly, for Bob we have
\beq p_B(b|x,y)=p_B(b).\label{eq:NBTSB2}  \eeq

Another interesting situation in this case is when one experiment, say Alice's, comes before Bob's, i.e. such that there is time for the system to pass from Alice to Bob, but not from Bob to Alice. Here the NBTS conditions are
\begin{align}
 p_A(a|x,y)&=p_A(a) \label{eq:NBTSA3} \\
 p_B(b|x,y)&=p_B(b|x)\label{eq:NBTSB2} 
\end{align}

as the state of Bob's system may now depend on Alice's input $x$ but the state of Alice's system cannot depend on $y$.

\item \textbf{Indefinite relative timing}:
In the other case we have no knowledge whatsoever about the relative timings of Alice and Bob.  It could be the case that the timing is well defined, but is unknown by Alice and Bob, or it varies from one run of the experiment to the next, or that the timing is undefined even in principle. The latter possibility may be very important, being relevant for situations such as quantum gravity, where a global time might not even exist.

In this case it is possible that the system that enters Alice's laboratory previously passed through Bob's laboratory. Thus Alice's measurement result $a$ could depend on $y$, since the system could have been transformed according to $y$ in the past. In other words, since  Bob could have affected Alice's system,  $p_A(a|y)$ could actually depend on $y$. Similarly, since it is also possible that the system that enters Bob's lab could have previously passed through Alice's lab, his probability may depend on Alice's input, i.e. $p_B(b|x)$ could depend on $x$. Hence in this situation the NBTS conditions are just the basic equations  \eqref{eq:NBTSA} and \eqref{eq:NBTSB} which describe that locally, in each lab, the probabilities cannot depend on the inputs received into that lab at a later time; no supplementary constraints apply. The NBTS conditions here are therefore weaker than those in the previous case.
\end{itemize}
Beyond these two extremal cases one could consider intermediate cases, for example, where Alice and Bob know the probabilities of the different time orderings.

%\footnote{One might also consider cases where all of the relevant events in Alice's labs do not have the same temporal relation with the all events in Bob's lab. However, all the matters for our arguments is whether Alice's input lies in the past lightcone of Bob's outcome or not, and vice versa, and all possible cases reduce to the three cases previously considered.}

In the subsequent sections, we explore some of the consequences of the NBTS constraints in the different scenarios, and their implications for particular theories. 

\section{Definite relative timing}
We start by exploring the case where there are well defined relative times between the laboratories. We will look at two situations: one where the experiments are in parallel (i.e. such that it is impossible that the same system passes through Alice's and Bob's labs, since there is not enough time for a system to travel from one lab to the other), and one where one party comes before the other (where will we take Alice to be first without loss of generality).

\subsection{No-backwards-in-time-signalling polytopes}

Our first goal is to characterise the set of correlations which can arise in these situations. We start with the situation where Alice's and Bob's experiments are in parallel. In this situation the NBTS conditions are
\begin{equation} \label{e:NBTS parallel}
\begin{split}
p_A(a|x,y) &= p_A(a)  \\
p_B(a|x,y) &= p_B(b)
\end{split}
\end{equation}

We will make use of the well known concepts and techniques used for analysing non-local correlations. In particular the geometrical tool of the `no-signalling polytope' can be used in our case too. 

For simplicity first consider the case when $a,b,x,y$ can each only take two values, $0$ and $1$.  The entire physical situation is then described by the 16 numbers $p(a,b|x,y)$. We can cast the situation in a geometrical form by considering a 16-dimensional space and associating to each physical situation a point $\PP$ whose coordinates are $\{ p(0,0|0,0),..., p(1,1|1,1)\}$. Since each coordinate is actually a probability, its values can only range between 0 and 1, meaning that the points $\PP$ describing physical situations live inside the 16-dimensional hypercube defined by $0\leq p(a,b|x,y)\leq 1$. The 4 probability normalisation relations $\sum_{a,b}p(a,b|x,y)=1$ for each pair $x,y$, impose further constraints, specifying hyperplanes on which the points $\PP$ must be situated. This means that the points $\PP$ can be only be situated in the 12-dimensional polytope obtained by intersecting the original 16-dimensional hypercube with the four  normalisation hyperplanes.  That is the space of all conceivable physical situations.

The no-backwards-in-time-signalling constraints  \eqref{e:NBTS parallel} further limit the space, specifying further hyperplanes on which $\PP$ must lie. The resulting `no-backwards-in-time-signalling polytope' is 6-dimensional and is the basic object we are interested in. In particular, one finds that this polytope has 18 vertices. Four of these vertices are  `deterministic' and given by 
\begin{equation}
p(a,b|x,y) = \begin{cases} 1 & \text{if } a= \alpha,\quad b =\beta \\
0 & \text{otherwise} \end{cases}
\end{equation}
where $\alpha, \beta \in \{0,1\}$. That is, for these vertices, $a$ and $b$ both take on constant deterministic values (equal to $\alpha$ and $\beta$ respectively). 
There are also  8 `PR-like' \cite{pr1994} vertices 
\begin{equation}
p(a,b|x,y) = \begin{cases} \tfrac{1}{2} & \text{if } a\oplus b =(x \oplus \gamma)(y\oplus \delta) \oplus \epsilon \\
0 & \text{otherwise} \end{cases}
\end{equation}
where $\gamma, \delta, \epsilon \in \{0,1\}$ and $\oplus$ denotes addition modulo 2. Formally these correlations look very similar to the `PR-box' which arises in the study of nonlocal correlations \cite{pr1994}. Note however that in the case of nonlocality, the inputs $x$ and $y$ are received before the outputs $a$ and $b$ are produced, whereas here $a$ and $b$ are produced before Alice and Bob receive the inputs $x$ and $y$.  

Finally, there are 6 `linear correlation' vertices given by
\begin{equation}\label{e:NBTS definite}
p(a,b|x,y) = \begin{cases} \tfrac{1}{2} & \text{if } a\oplus b =\alpha x \oplus \beta y \oplus \delta\\
0 & \text{otherwise} \end{cases}
\end{equation}
where $\alpha, \beta, \delta \in \{0,1\}$ and $\alpha$ and $\beta$ cannot be both simultaneously 0.

For both the PR-like vertices and the linear correlation vertices, we see interesting structure: the sum (modulo 2) of the results, $a \oplus b$, has a non-trivial dependence on the later inputs $x$ and $y$. That is, the correlations between the results obtained in the two labs are affected by the later inputs. Looking only at Alice's or Bob's results however, the NBTS conditions are satisfied, and there is no dependence on either $x$ or $y$. This is exactly like in the case of nonlocality, where it is precisely in the correlations that nonlocality arises.  

Moving on to the situation where Alice's experiment takes place before Bob's, then we find, again in the simple case of $a, b, x, y \in \{0,1\}$, that the corresponding NBTS polytope is now one dimension larger compared to the previous case, being 7-dimensional, and has 20 vertices. Here we find that there are 8 deterministic vertices given by 
\begin{equation}
p(a,b|x,y) = \begin{cases} 1 & \text{if } a= \alpha,\quad b =\beta x \oplus \gamma \\
0 & \text{otherwise} \end{cases}
\end{equation}
where $\alpha, \beta, \gamma \in \{0,1\}$. That is, Alice's outcome is now identical to before (being determinstic and constant) but now Bob's output is still determinstic, but no longer constant in general, but rather a function of Alice's input $x$.

The same 8 PR-like correlations from the previous case remain vertices,   
\begin{equation}
p(a,b|x,y) = \begin{cases} \tfrac{1}{2} & \text{if } a\oplus b =(x \oplus \gamma)(y\oplus \delta) \oplus \epsilon \\
0 & \text{otherwise} \end{cases}
\end{equation}
where $\gamma, \delta, \epsilon \in \{0,1\}$. In contrast to before, there are now only 4 linear correlation type vertices, given by
\begin{equation}\label{e: A before B}
p(a,b|x,y) = \begin{cases} \tfrac{1}{2} & \text{if } a\oplus b =y \oplus \alpha x \oplus \beta\\
0 & \text{otherwise} \end{cases}
\end{equation}
where $\alpha, \beta, \in \{0,1\}$, i.e only those which have a dependence on $y$ remain. This is due to the fact that by taking convex combinations of the 4 new deterministic vertices that have an $x$ dependence, it is possible to produce the two linear-correlations which were previously vertices. 

\subsection{Classical polytopes}\label{sec:classical} 
Our second goal is to characterise the set of correlations that could arise in a classical setting, such that the system that enters Alice's and Bob's lab is a classical variable, which we will denote by $\lambda$. This is similar to the classical models used in the context of nonlocality, where $\lambda$ is traditionally referred to as a `hidden variable', and it used by Alice and Bob to produce their correlations.

We will start with the parallel case, where Alice's and Bob's actions overlap in time, and denote this by $A|B$. Here, a copy of $\lambda$ enters both labs, with probability density $\rho(\lambda)$. Alice and Bob can then generate $a$ and $b$ depending upon $\lambda$, according to probability distributions $p_A(a|\lambda)$ and $p_B(b|\lambda)$ respectively. Since both outcomes must be given before $x$ and $y$ are input, $a$ and $b$ cannot depend on either $x$ or $y$. The joint probability distributions that can be generated are thus
\beq p^{A | B} (a,b|x,y)=  \int d\lambda \rho(\lambda) p_A (a|\lambda) p_B(b|\lambda).  \eeq
We note that this form is sufficient to generate any joint probability distribution which does not depend upon $x$ or $y$, i.e.
\begin{equation}
p^{A | B} (a,b|x,y)= p_{AB}(a,b).
\end{equation}
Imposing this constraint, along with the normalisation and positivity constraints, leads to the classical polytope. For the case where $a,b,x,y$ can each only take two values, $0$ and $1$ we find that the classical polytope can be readily found, is only 3 dimensional, with 4 vertices given by the deterministic vertices of the NBTS polytope 
\begin{equation}
p(a,b|x,y) = \begin{cases} 1 & \text{if } a= \alpha,\quad b =\beta \\
0 & \text{otherwise} \end{cases}
\end{equation}

\begin{figure}[t!]
	\begin{center}
\includegraphics[width=1\columnwidth]{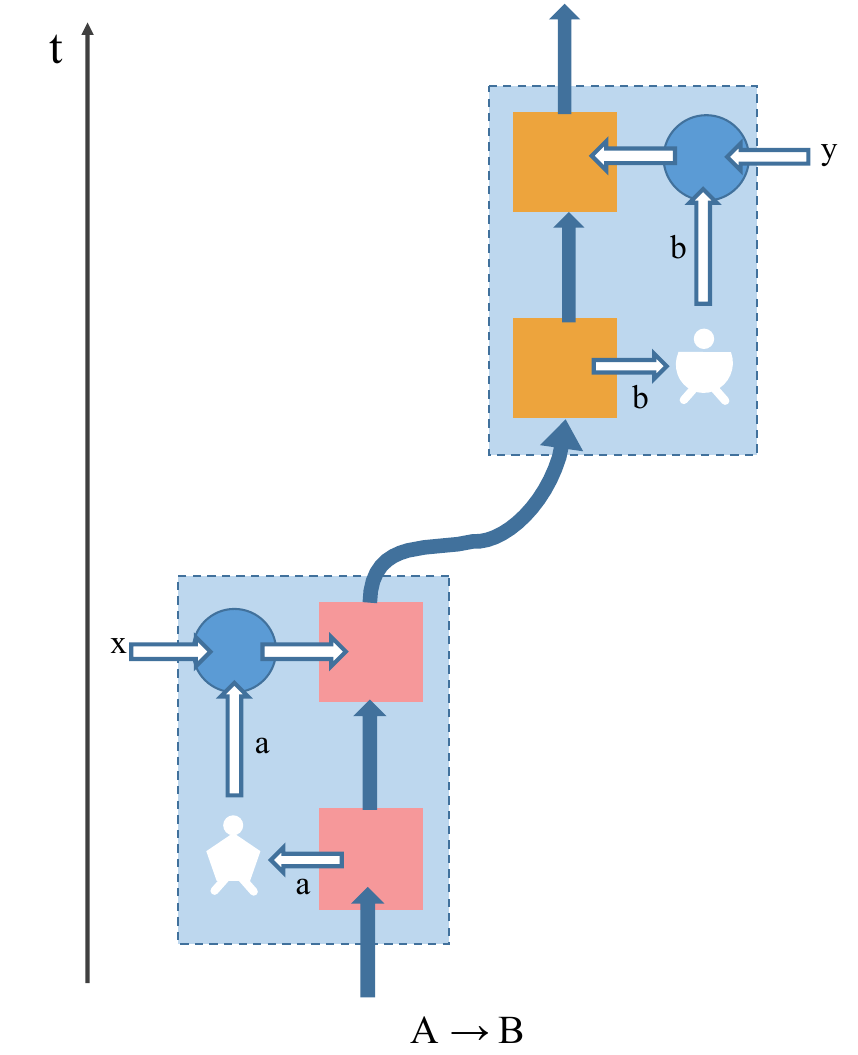}
\caption{The classical situation $A \to B$, in which Alice's operations occur before Bob's. It is therefore possible for a system to pass from Alice's lab into Bob's, and thus for the outcome $b$ to depend upon Alice's input $x$. However, note that in this case, all outcome probabilities are independent of the transformation labelled $y$, as it occurs after all measurements have been made. \label{f:classical}}
	\end{center}
\end{figure} 
Thus, we see that the 14 PR-like and linear-correlation vertices of the NBTS polytope are non-classical, and cannot be generated by such a classical model. These vertices are thus akin to the nonlocal vertices of the non-signalling polytope. 

In the case of nonlocality, once the local vertices of the non-signalling polytope are identified, which leads to the local polytope, the next step is usually to find all of the linear inequalities that separate it from the rest of the non-signalling polytope. These are precisely the Bell inequalities. In the present case, the situation is completely different from the case of the no-signalling polytope. In particular, whereas in the case of nonlocality the dimension of the local polytope is the same as the dimension of the non-signalling polytope, now we find that the classical polytope is of lower dimension than the dimension of the NBTS polytope (being 3 dimensional, rather than 6 in the setting considered).

The additional equalities that are obeyed, that reduce the dimension of the classical polytope relative to the NBTS polytope are given by
\begin{equation}
p(a,b|0,0) = p(a,b|x,y)
\end{equation} 
for all $x$,$y$. That is, these equalities encode directly that $a$ and $b$ follow a fixed probability distribution. On the other hand, we find that the only inequalities that need to be satisfied by the classical probability distributions are the positivity inequalities, demanding that all probabilities are non-negative. Thus, there are no new Bell-type inequalities, and only new equality constraints. This is the opposite of the nonlocality case, where the local polytope does not satisfy any new equality constraints but does satisfy new Bell-inequality constraints.

We now move on to the second situation, in which Alice's actions occur entirely before Bob's, which we denote $A \rightarrow B$. Now, the classical system $\lambda$ enters Alice's laboratory with probability density $\rho(\lambda)$, which she uses to produce $a$ according to $p_A(a|\lambda)$. She then receives $x$, and uses this to transform $\lambda$ into $\mu$, according to some probability density $\rho'(\mu|a,x,\lambda)$. She then sends $\mu$ to Bob, who will use it to produce $b$ according to $p_B(b|\mu)$ (see Fig.~\ref{f:classical}). The achievable probability distributions via this time ordering are therefore 
\begin{multline} 
p^{A \rightarrow B} (a,b|x,y)= \\\iint d\lambda d\mu \rho(\lambda)\rho'(\mu|\lambda,a,x) p_A(a|\lambda)   p_B(b|\mu), 
\end{multline}
Note that, without loss of generality, we can assume that $\mu = (\lambda,a,x)$, i.e. the system that leaves Alice's lab is just the combination of all the information that entered it, along with her measurement outcome, in which case we can equivalently find 
\begin{multline} 
p^{A \rightarrow B} (a,b|x,y)= \int d\lambda \rho(\lambda)p_A(a|\lambda)   p_B(b|a,x,\lambda), 
\end{multline}
Defining 
$
p_A(a) = \int d\lambda \rho(\lambda) p_A(a|\lambda)
$  and 
$p_B(b|a,x) = \int d\lambda \rho(\lambda) p_A(a|\lambda) p_B(b|a,x,\lambda)/p_A(a)  $
%
%$p_A(a) = \int d\lambda \rho(\lambda) p_A(a|\lambda)$ 
% and $p_B(b|a,x) = \int d\lambda \rho(\lambda) p_A(a|\lambda) p_B(b|a,x,\lambda)/p_A(a)$, 
 we arrive at the simplest form\footnote{Note that if $p_A(a) = 0$, for some outcomes $a$, then we can define $p_B(b|a,x)$ arbitrarily for these outcomes (to avoid dividing by zero), as they never occur.}
\begin{equation}
p^{A \rightarrow B} (a,b|x,y)= p_A(a)p_B(b|a,x).
\end{equation}
This is the form of the most general classical probability distribution that is consistent with the fixed timing of Alice before Bob. 

For the case considered previously, with each party having binary inputs and outputs, the classical polytope is $5$ dimensional, with $8$ vertices given again by the deterministic vertices of the corresponding NBTS polytope 
\begin{equation}
p(a,b|x,y) = \begin{cases} 1 & \text{if } a= \alpha,\quad b =\beta x \oplus \gamma \\
0 & \text{otherwise} \end{cases}
\end{equation}

We again find that the only inequality constraints satisfied by the classical polytope are the positivity inequalities (hence there are again no Bell-type inequalities). The new equality constraints that are obeyed are found to be
\begin{equation}
p(a,b|x,0) = p(a,b|x,y)
\end{equation} 
for all $x$, $y$. This again directly encodes that $a$ must be constant (but now $b$ is allowed to be a function of $x$). 

Thus, in both the case of parallel timing, and in sequential timing we find a similar structure: the NBTS polytopes contain both classical and non-classical vertices, and the classical polytope is of lower dimension that the NBTS polytope. We will explore some of the consequences of these findings in the next two sections.
%Crucially, everything else outside the classical polytope {\bf (??? is this true)} presents backwards in time influence without backwards in time signalling.

\subsection{Backward in time influence without backward in time signalling}
A very interesting phenomenon is the possibility of the future affecting the past, without backwards in time signalling. This possibility appears clearly in situations with well defined relative timing between Alice and Bob. 

Suppose, for example, that Alice and Bob's experiments are in parallel. We can then envisage that the results of Alice and Bob's measurements, which occur on Monday, may depend on the external inputs $x$ and $y$ that they receive on Tuesday. However, if only the correlations between $a$ and $b$ are affected (not the marginals), this dependence can only be observed on Wednesday when Alice and Bob emerge from their labs and compare their results to check their correlations. 

Mathematically, this is the case when $p_A(a|x,y)=p_A(a)$, $p_B(b|x,y)=p_B(b)$, ensuring no backwards in time signalling can be observed by Alice and Bob while in their labs,  but  $p(a,b| x,y)\neq p(a,b)$, i.e. the correlations depend on $x$ and $y$ that both occur after obtaining $a$ and $b$. In particular, consider the NBTS correlations given by \eqref{e:NBTS definite} with $\alpha = \beta = \delta = 1$, i.e. such that $a \oplus b = x \oplus y \oplus 1$, but $a$ and $b$ are individually uniformly random. This correlation has the above properties, and hence in this case the future affects the past, without backwards in time signalling.

Similarly, when Alice's experiment takes place before Bob's, a similar situation arises when $p_A(a|x,y)=p_A(a)$, $p_B(b|x,y)=p_B(b|x)$ but $p(a,b| x,y)\neq p(a,b)$ as in \eqref{e: A before B}.

\subsection{Avoiding potential paradoxes}\label{s:avoiding paradoxes}
Imagine now that Alice and Bob carry out some procedures in their labs, which leads to the correlations $p(a,b|x,y)$ being produced. An interesting question one can ask  is what would happen if Alice changed her procecure to completely ignore the externally generated $x$, and instead generate internally an $X$ (which may depend upon $a$) which she then uses in place of $x$ to determine which transformation to make on the system. Apart from switching from $x$ to $X$, she otherwise follows the same procedure as  before. Similarly, Bob could internally generate a $Y$ (depending on $b$), and use this in place of $y$. If we assume that the physics does not distinguish how the choice of transformation was made, and therefore $p(a,b |x,y)=p(a,b |X,Y)$, we could, in some cases run into paradoxes. For example, given the correlation before, such that $a \oplus b = x \oplus y \oplus 1$, if Alice now chooses $X = a$ and Bob chooses $Y = b$, then demanding $a \oplus b = X \oplus Y \oplus 1$ would lead to contradiction.

The issue however is that ignoring $x$ and $y$ means that Alice and Bob are performing different procedures in their labs from what they were originally performing, resulting in physically different states of the lab.  As a consequence $p(a,b|x,y)$ does not need to equal $p(a,b|X,Y)$, and paradoxes can be avoided. The precise mechanism for this will depend on the details of the underlying model that determines how the correlations arise in the first place. In Sec.~\ref{s:two-time states}, we consider a particular model for generating correlations (two-time quantum states). However we find that they are only able to generate classical correlations, not ones that would potentially lead to paradoxes. It is an interesting direction for future research to develop models which lead to nonclassical correlations and to explore how paradoxes are avoided in such models.

%In cases with indefinite relative time ordering the situation is far more subtle and needs further analysis.

\section{Indefinite relative timing}
We now move on to the task of exploring the case where we do not assume a fixed relative time between the laboratories. As mentioned already, it could be the case that the timing is well defined, but is unknown to Alice and Bob, or that it varies from one experimental run to another; it could even be the case that the timing is undefined in principle. The latter possibility may be very important in quantum gravity, where a global time might not even exist.

\subsection{No-backwards-in-time-signalling polytope}

Our first goal is again to characterise the set of correlations which can arise in this situation. 

We note in the present scenario with indefinite relative timing, the NBTS conditions \eqref{eq:NBTSA} and \eqref{eq:NBTSB} are almost identical to the no-signalling constraints from the study of non-local correlations. The NBTS conditions are
\begin{equation}
\begin{split}
p_A(a|x,y)&=p_A(a|x)\\ 
p_B(b|x,y)&=p_B(b|y),
\end{split}
\end{equation}
where we see that the roles of $x$ and $y$ have been reversed, compared with their counterparts from nonlocality. As a consequence, the NBTS polytope for indefinite relative timing is closely related to the no-signalling polytope. In particular, the vertices of each are in one-to-one correspondence, except $x$ and $y$ are swapped. 

Thus, the NBTS polytope has 24 vertices. 16 of these correspond to deterministic probability distributions
\begin{equation}
p(a,b|x,y) = \begin{cases} 1 & \text{if } a=\mu y \oplus \alpha,\quad b = \nu x\oplus\beta \\
0 & \text{otherwise} \end{cases}
\end{equation}
where $\alpha$, $\beta$, $\mu, \nu \in \{0,1\}$, where $a$ and $b$ take on deterministic values as a function of $y$ and $x$ respectively. 

The remaining 8 vertices are again `PR-like' \cite{pr1994} correlations, given by 
\begin{equation} \label{eq:prcase}
p(a,b|x,y) = \begin{cases} \tfrac{1}{2} & \text{if } a\oplus b =(x \oplus \gamma)(y\oplus \delta) \oplus \epsilon \\
0 & \text{otherwise} \end{cases}
\end{equation}
where $\gamma, \delta, \epsilon \in \{0,1\}$. 
As in the previous case, we again see non-trivial dependence of the sum $a \oplus b$ of measurement results on the future inputs into the labs. 

\subsection{Classical polytope} \label{sec:classical}

\begin{figure*}
\begin{center}
\includegraphics[width= 1.55\columnwidth]{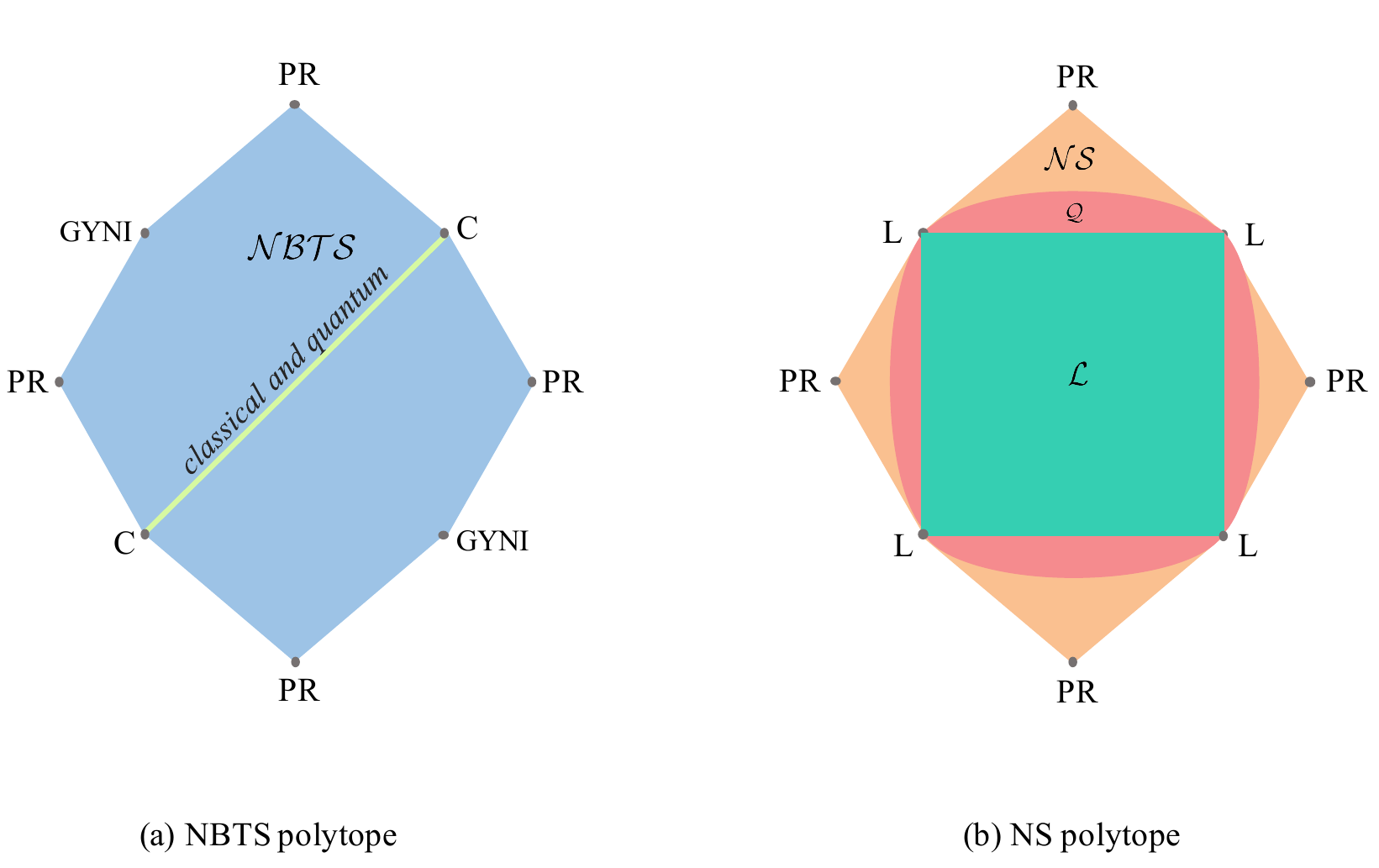}
\caption{A comparison of the no-backwards-in-time signaling polytope (left) with the no-signalling polytope (right) for the case of indefinite relative timing between the labs. (a) In the NBTS case, the classical polytope coincides with the quantum polytope, and is of lower dimension, depicted by the solid diagonal line. Vertices labelled $GYNI$ correspond to vertices which always win `guess-your-neighbours-input' type games \cite{GYNI}. $PR$ vertices correspond to Popescu-Rohrlich type correlations \cite{pr1994}. $C$ corresponds to deterministic vertices, where at most one party's output depends on the other's input. (b) In the NS case, the local polytope $\mathcal{L}$ (with vertices denoted L) is full-dimensional in the space of non-signalling correlations. The quantum set, $\mathcal{Q}$, lies between the local and non-signalling polytopes. \label{f:polytope}  }
\end{center}
\end{figure*}

We are now interested once again in identifying the subset of the NBTS polytope which can be generated in a classical world, i.e.~ where both the systems and the time-ordering are entirely classical, which will allow us to identify those correlations which are non-classical.  

Classically, in any run of the experiment the time ordering between Alice and Bob is well defined. There are three possibilities --- either Alice's actions occur entirely before Bob's, which we denote $A \rightarrow B$, Bob's actions occur entirely before Alice's, $A \leftarrow B$, or their actions overlap in time, $A|B$. Crucially, as we are now considering indefinite relative timing, we imagine that the time ordering is chosen at random in each run, and is unknown to Alice and Bob. 

In the first case, in which Alice's actions occur entirely before Bob's, the situation is identical to the definite relative timing of Sec.~\ref{sec:classical}, and hence the most general distributions are of the form 
\beq p^{A \rightarrow B} (a,b|x,y)= p_A(a) p_B(b|a,x). \eeq
where $p_A(a)$ and $p_B(b|a,x)$ are arbitrary probability distributions. Similarly, when Bob's actions occur before Alice, we obtain 
\beq p^{A \leftarrow B} (a,b|x,y)= p'_A(a| b,y) p'_B(b). \eeq
Finally, when their actions overlap in time,  we have as before
\beq p^{A | B} (a,b|x,y)=  p_{AB}(a,b)  \eeq
However, note that the situation $A|B$ is strictly weaker than the other two cases, as the outputs are independent of the inputs. Any correlations of this form could be generated by either of the other two cases and can be absorbed into them without loss of generality.
All classically achievable probability distributions can therefore be written as 
\begin{align}
\begin{split}
  p(a,b|x,y)=q p_A&(a) p_B(b|a,x) \\
  &+ (1-q)  
	p'_A(a| b,y) 
		p'_B(b) 
\end{split}
\end{align}
where $q\in[0,1]$ denotes the probability of the time-ordering $A \rightarrow B$, and $p_A(a), p_B(b|a,x), p'_A(a| b,y) $ and $p'_B(b)$ are valid probability distributions. Such probability distributions define the `classical polytope' for a given number of inputs and outputs. 

For the case considered previously, with each party having binary inputs and outputs, the `classical polytope' is once again straightforward to construct. It is 7 dimensional and has 12 vertices given by 
\begin{equation}
p(a,b|x,y) = \begin{cases} 1 & \text{if } a=\mu y \oplus \alpha,\quad b = \nu x\oplus\beta \\
0 & \text{otherwise} \end{cases}
\end{equation}
with $\alpha$, $\beta$, $\mu, \nu \in \{0,1\}$ and  $\mu$ and $\nu$  not simultaneously equal to 1. It corresponds to the subset of deterministic vertices of the NBTS polytope which do not depend on both $x$ and $y$. 

Once again there are no new inequalities satisfied by classical correlations beyond positivity constraints. The new equality constraints that must be satisfied are found to be
\beq  p(a,b|x,y)+p(a,b|x',y')=p(a,b|x,y')+p(a,b|x',y)\label{eq:classicalequality} \eeq
for all $a, b, x,y,x'$ and $y'$.  Although at first sight these conditions are not as straightforward to interpret as in the previous case of definite relative timing, they can still be easily understood by noting that each classical situation can be expressed as a mixture of ${A \rightarrow B}$ and $ {A \leftarrow B}$ cases. In the former case, $p^{A \rightarrow B} (a,b|x,y)$ is independent of $y$ and thus 
\begin{align}
\begin{split}
 p(a,b|x,y) &= p(a,b|x,y')\\
&\text{and} \\
 p(a,b|x',y) &= p(a,b|x',y') 
 \end{split}
 \end{align}
which ensures that \eqref{eq:classicalequality} is satisfied.  In the latter case, $p^{A \leftarrow B} (a,b|x,y)$ is independent of $x$ and thus
\begin{align}
\begin{split}
p(a,b|x,y) &= p(a,b|x',y)\\
&\text{and} \\
p(a,b|x,y') &= p(a,b|x',y')
\end{split}
\end{align}
which also satisfies \eqref{eq:classicalequality}. As both cases independently satisfy  \eqref{eq:classicalequality}, any mixture of them also will. 

The new equalities given by \eqref{eq:classicalequality} are not all independent of the NBTS conditions. In particular, for the  case considered earlier in which each party has binary inputs and outputs, the single equality  
\beq  p(0,0|0,0)+p(0,0|1,1)=p(0,0|0,1)+p(0,0|1,0)\label{eq:singleclassicalequality} \eeq
can be combined with the NBTS conditions to generate all of the other cases. For more details see Appendix~\ref{ap:classical}.

Thus the classical polytope is obtained precisely by taking the NBTS polytope and imposing the additional equality \eqref{eq:singleclassicalequality}, which is why its dimension is one smaller. 

We show in Appendix~\ref{ap:classical2} that when considering two parties with any number of inputs and outputs, the classical polytope for indefinite relative timing can be obtained by intersecting the NBTS polytope with the equalities given in \eqref{eq:classicalequality}. That is, the conditions \eqref{eq:classicalequality} completely characterise classicality in the two-party scenario --- they constitute the necessary and sufficient conditions that need to be satisfied (in addition to the NBTS, normalisation and positivity conditions), in order for a distribution to be classical. This will turn out to be important in the next section, when we look at using two-time quantum states as a concrete model for studying no backwards in time signalling. 

%For cases with more inputs and outputs,  we conjecture that the classical polytope can be obtained in the same way, as the intersection of the NBTS polytope, and the additional equalities given by \eqref{eq:classicalequality}. We have verified this explicitly for several cases with small inputs and outputs (given in Table~\ref{table1} in Appendix~\ref{ap:classical2}), but a general proof of this remains an interesting open problem. 

\section{Two-time quantum states}\label{s:two-time states}

In the standard quantum formalism, the timing of all measurements are fixed. Considering such cases, or mixtures of them, we obtain precisely the same polytope of correlations as in the classical case. 

However, as discovered by Aharonov, Bergmann and Lebowitz \cite{ABL}, and developed further in subsequent works \cite{Aharonov1991,Aharonov,multi,Ralph2014}, in quantum mechanics systems can have imposed on them a final state in addition to and independently from the initial state. The initial and final states together determine what happens at intermediate times. We call such situations `two-time states'. Nature could provide the final boundary condition.  Such situations have never been observed, but they are theoretically possible. In fact there are a few proposals for the possible existence of such situations: as a final state of the Universe or a final state at the singularity of a black hole \cite{AhaPeronal,AhaRoh05,Aharonov,Hartle,Gell-Mann,maldacena}.

Note that the fundamental postselection described above can be simulated easily experimentally. For example,  by performing a measurement at the final time, and considering only those cases which yield a subset of the possible results.

Two-time states in general allow for backwards in time signalling between two intermediate times. It is interesting to understand whether there could be non-trivial subclasses of two-time states that do not lead to backward in time signalling in the situations considered in the previous section, namely when there is definite or indefinite relative timing. Such a subclass would then provide a model with which to study the general results obtained previously. 

In the case of parallel experiments with definite relative timing, in the appendix we show that the only two-time states that do not lead to signalling are those without post-selection, or with trivial post-selection\footnote{i.e. post-selection on the maximally mixed state.}. Thus in this case, it is not possible to find a non-trivial class of states.

For the case of definite relative timing with Alice's experiments coming before Bob's, we show in the appendix that there is a non-trivial class of two-time states that do not lead to backward-in-time signalling. These correspond to states where for Bob there is no post-selection, or a trivial post-selection taking place. However, Alice is indeed allowed to perform a non-trivial post-selection in general.  

Finally, for the case of indefinite relative timing, we show in Appendix~\ref{ap:two-time} (and sketch below) that there again exists a non-trivial subclass of two-time states that does not allow for backwards in time signalling. 

These particular two-time states have a special property: the probability of succeeding to prepare such a state in an experimental simulation is independent of the measurements performed on the system at intermediate times between pre- and post-selection, as long as the measurements are of local observables for Alice and Bob. As a consequence, the probability of obtaining a given outcome for an intermediate time measurement depends linearly on the measured operators, a property that is not valid in general for two-time states.

There is an important observation to be made here: The correlations that arise in both cases from the corresponding subset of two-time states obey the classicality conditions which are obeyed by classical correlations. Hence, they can only produce classical correlations.

A detailed proof of this result is given in Appendix~\ref{ap:two-time}, using a formalism developed specifically for analysing  pre- and post-selected situations. However, we sketch the key ideas below. 

We first consider the most general pre- and post-selected quantum states for a single party that  satisfy NBTS. Consider that you prepare an arbitrary quantum state of a system and ancilla, and pass the system into Alice's lab. Alice then performs a quantum measurement, obtaining output $a$, followed by a transformation labelled by $x$. Finally, she outputs the resulting system, and you perform a post-selected measurement on the system and ancilla. We can characterise the entire procedure outside Alice's lab by a two-time state on her input and output spaces. We will say that this state satisfies NBTS if $ p(a|x)=p(a)$ for all choices of measurement and transformation by Alice. 

By considering a sufficiently large set of measurements and transformations\footnote{in particular, destructive measurements in a selection of bases, followed by preparation of the $\ket{0}$ state, and a range of unitary transformations}, we show in Appendix~\ref{ap:two-time} that a  single party two-time state  satisfies NBTS if and only if it corresponds to a case without post-selection, or with trivial post-selection.

In the case of two parties, the NBTS conditions state that when we sum over Bob's output $b$, then Alice's marginal probability distribution (which may in general  depend on $y$) must satisfy the single-party NBTS conditions. Given the above result, it follows that from Alice's perspective there must either be no post-selection, or a trivial post-selection (where the probability of success is independent of $x$). Following the same argument with the parties reversed, we can show that the probability of a successful  post-selection must also be independent of $y$. Hence, the probability of a successful post-selection must be independent of both $x$ and $y$. This is sufficient to imply that the situation can be represented by a linear two-time state. To prove the converse, that any linear two-time state satisfies NBTS, we can use the theorem from \cite{Silva2017}. 

Note that when simulating a linear two-time state of two parties via experimental post-selection, the probability of success is a constant for any local operations of Alice and Bob. However, if Alice and Bob were to combine their laboratories and perform some joint measurement on their systems, then this could in principle affect the post-selection probability. Hence, such states may involve non-trivial post-selection at a global level, even though the local effects appear trivial.   

\section{Process Matrices}

Recently, a  framework for  correlations has been investigated which does not assume a global causal order, but only the local validity of quantum theory --- leading to the discovery of correlations with indefinite causal order \cite{Oreshkov2012}.  The  key object in this formalism is the process matrix, which captures the connection between Alice's and Bob's laboratories. The setup considered is similar to the one presented in this paper, except for the crucial difference that the time ordering of the inputs and outputs in each lab is reversed. In the process matrix formalism Alice first receives an input $x$, then performs a measurement depending on this to generate her output $a$, whereas in the setup considered in this paper she first performs a fixed measurement to produce $a$, and then receives an input $x$ and performs a transformation depending on this. The process matrix formalism includes all quantum processes with definite time ordering ($A \rightarrow B, A \leftarrow B, A|B$) but also includes cases which cannot be explained by any mixture of such processes. 

A physical mechanism for generating such indefinite causal correlations was originally left open. However, it was recently shown \cite{Silva2017} (see also \cite{Oreshkov2016}) that any process matrix can be simulated by  quantum theory with post-selection. Furthermore, the set of two-time  quantum states corresponding to valid process matrices are precisely the linear two time states described in the previous section. In the context of pre- and post- selected states,  linearity seems a somewhat arbitrary and technical restriction.  The results presented in the previous section provide a physical motivation for this set, by showing that it contains precisely those two-time quantum states which satisfy the NBTS conditions of the indefinite relative timing scenario. It also follows  that any situation described by a process matrix cannot yield backwards-in-time signalling, in the sense considered here. Furthermore, although in general process matrices can lead to non-classical correlations \cite{Oreshkov2012}, for the situations we consider here, surprisingly they can only generate classical correlations, i.e.~ correlations in the classical polytope of Sec.~\ref{sec:classical}. This  generalises an earlier result (considered in a different context) \cite{Baumann2016}, which implies that Alice and Bob can only generate classical correlations when each performs a fixed basis measurement followed by a variable transformation on a process matrix. Our results show that any fixed measurement (including  POVMs or  projective measurements involving projectors of any rank) followed by a variable transformation will also lead to classical correlations between Alice and Bob.

\section{Discussion}
We have presented a theory-independent definition of no-backwards-in-time-signalling, that  is a temporal analogue of the no-signalling conditions that lie at the heart of research into non-locality. What we discovered is that in probabilistic theories it is theoretically possible to have situations (such as in \eqref{eq:prcase}) in which the future demonstrably affects the past, but in such a way that the effect can only be discovered later,  thereby avoiding paradoxes such as  killing one's own grandfather. Such situations have not yet been observed, so their existence is purely speculative at the moment. However  it is nevertheless  instructive to understand the full scope of possible natural laws  which avoid these paradoxes.

In a scenario with two parties, we split our study into three cases, which distinguish the prior knowledge we have about the relative time order between the two labs. In the first case, where we have no knowledge (indefinite relative timing), the NBTS polytope is isomorphic to the  no-signalling polytope of the same scenario. However, the set of correlations which can be achieved classically (which obey standard `forwards-in-time' causality) differ from the analogous local polytope, obeying additional equalities. This means that they lie in a lower dimensional subset of the full NBTS polytope (i.e. of relative measure zero).

In the second case we know that the actions of Alice and Bob happen in  parallel (definite parallel timing), which gives additional constraints which restrict the scenario above. The corresponding NBTS and classical polytopes are of lower dimension than for the case with no knowledge about the relative timings.

Finally, in the third case we have knowledge about the relative timing between the labs, for example, that Alice's experiment took place before Bob's (definite sequential timing). Here, the NBTS and classical polytopes are of intermediate dimension between the first and second cases since the linear constraints represent a relaxation of the most restrictive case of parallel timing. These results are summarised in Table \ref{tab:summary}.

\begin{table}[t!]
\hspace{-0.46cm}
\footnotesize
\begin{tabular}{cllllr}
\textbf{Order}      &   \multicolumn{2}{c}{\textbf{Constraints} }   & \textbf{Polytope}  \\ &&& (2-input,\\&&& 2-output) \\ \cline{1-4}
\multirow{2}{*}{\textbf{\begin{tabular}[c]{@{}c@{}}Indefinite \\ relative \\timing \end{tabular}} }                                              
\T\T &NBTS&	$\begin{aligned} \sum_a p(ab|xy)&= p_B(b|x) \\
	\sum_b p(ab|xy)&= p_A(a|y) \end{aligned}$ & 
	\begin{tabular}[c]{@{}c@{}}$8$-$d$\\ $24$ vertices \end{tabular} 
		&& \\ \cline{2-4} \B &
				\begin{tabular}[c]{@{}l@{}}Classical\\ and quant.\end{tabular} 
					&	$\begin{aligned} p(ab|xy) = q p^{A\rightarrow B} \\ + (1-q)p^{B\rightarrow A}  \end{aligned}$
& \begin{tabular}[c]{@{}c@{}}$7$-$d$\\ $12$ vertices \end{tabular}      \\\cline{1-4}
\multirow{2}{*}{\textbf{\begin{tabular}[c]{@{}c@{}}Definite \\ parallel \\timing \end{tabular}} }                                                            
\T\T &NBTS&	$\begin{aligned} \sum_a p(ab|xy)&= p_B(b) \\
	\sum_b p(ab|xy)&= p_A(a) \end{aligned}$ &  \begin{tabular}[c]{@{}c@{}}$6$-$d$\\ $18$ vertices \end{tabular} && \\ \cline{2-4}
\B &\begin{tabular}[c]{@{}l@{}}Classical\\ and quant.\end{tabular}&	$\begin{aligned} p(ab|xy) = p_A(a) p_B(b)  \end{aligned}$
				&  \begin{tabular}[c]{@{}c@{}}$3$-$d$\\ $4$ vertices \end{tabular}                         \\\cline{1-4}
\multirow{2}{*}{\textbf{\begin{tabular}[c]{@{}c@{}}Definite \\ sequential\\ timing\\ $A\rightarrow B$\end{tabular}} }                                                      
\T\T &NBTS &	$\begin{aligned} \sum_a p(ab|xy)&= p_B(b|x) \\
	\sum_b p(ab|xy)&= p_A(a) \end{aligned}$ & \begin{tabular}[c]{@{}c@{}}$7$-$d$\\ $20$ vertices \end{tabular}   && \\ \cline{2-4}
 \B &\begin{tabular}[c]{@{}l@{}}Classical\\ and quant.\end{tabular}            &	$\begin{aligned} p(ab|xy) = q p^{A\rightarrow B} \\
+ (1-q)p^{B\rightarrow A}  \end{aligned}$
				& \begin{tabular}[c]{@{}c@{}}$5$-$d$\\ $8$ vertices \end{tabular}  \\\cline{1-4}
\end{tabular}
\caption[]{Summary of results. According to the knowledge about the relative timing between laboratories, different linear constraints apply and yield different polytopes for the spaces of joint correlations $\{p(ab|xy)\}$ between Alice and Bob. The constraints hold for an arbitrary number of inputs and outputs, while the polytopes described are for the specific case where $a, b,x,y \in \{0, 1\}$.}
\label{tab:summary}
\end{table}

Despite the mathematical similarity between the NBTS polytope and the no-signalling polytope in the case of indefinite timing, it is important to note that the physics of the two cases is very different.  In the case of non-local boxes, Alice and Bob are outside the boxes. They then use the boxes as resources, since whatever they do does not change the way in which the boxes act. On the other hand, in the NBTS scenario, we consider closed laboratories with Alice and Bob inside (and part of) their respective laboratory. Crucially, their actions can, and do, modify the correlations obtained within their laboratories (see discussion in Sec.~\ref{s:avoiding paradoxes}). Hence we cannot think of NBTS correlations as a resource in the same way as NL boxes.  %In particular, if instead of using the independently generated external inputs $x$ and $y$ to determine which transformations to perform on their systems (as we have considered throughout this work), Alice and Bob attempt to `close a time loop' by using their earlier measurement results to determine their transformations, they may create different time-flow structures inside their labs, so the correlations that arise are no longer the same as if they had used the external inputs. 

In the context of two-party two-time quantum states with indefinite relative timing, the NBTS condition exactly characterises the special set of cases corresponding to process matrices \cite{Oreshkov2012}. Furthermore,  the correlations achievable by such states are identical to those achievable classically\footnote{Note that the correlations are classical for the situation considered here, where Alice and Bob's measurements occur before their transformations. However, process matrices do lead to non-classical correlations in the situation where Alice and Bob perform transformations based upon their inputs.}. It has subsequently been shown \cite{purves2019} that for three or more parties, the NBTS conditions still characterise the linear two-time states (which also remain equivalent to process matrices) but now they can produce non-classical correlations. Thus it is a peculiarity of the two-party linear two-time states that they only yield classical correlations.

For the case of indefinite relative timing, the NBTS conditions can be generalised straightforwardly to multiple parties. Given $N$ parties, with inputs $\textbf{x}=\{x_1, x_2, \ldots x_N\}$ and outputs $\textbf{a}=\{a_1, a_2, \ldots a_N\}$, we demand that each party's marginal probability distribution is independent of their input. i.e. 
\begin{align}  \sum_{a_2, a_3, \dots a_N} P(\textbf{a} |\textbf{x}) \nonumber = P_A(a_1 | x_2, \ldots, x_N) 
\end{align} 
and similarly for the other parties. This ensures that each party individually does not perceive backwards in time signalling. Note that this is different from the usual multi-party no-signalling conditions, where we only sum over one party's input (e.g. $\sum_{a_1}P(\textbf{a} |\textbf{x})$ is independent of $x_1$). Hence, in general, the multi-party polytopes in the two cases will not be isomorphic. 
Finally, the multipartite cases of definite relative timings would present further constraints to the one above. The extremal case of when everyone's actions happen in parallel generalises trivially as
\begin{align}
 \sum_{a_2, a_3, \dots a_N}  P(\textbf{a} |\textbf{x}) =  P_A(a_1) 
 \,,
\end{align}
and analogously for the other parties. The other cases of definite relative timing are more subtle, since  they could contain some subset of the parties acting in parallel and then signalling to the others. For example in the case where Alice's actions come before Bob and Charlie's (who act in parallel), whose actions come before Dave, denoted $A\rightarrow B|C\rightarrow D$. It would be interesting to explore these cases further and find the general classes of behaviours. 
\acknowledgments YG acknowledges funding from the Austrian Science Fund (FWF) through the project P 31339-N27,  the Zukunftskolleg ZK03, the START project Y879-N27, the joint Czech-Austrian project Multi-QUEST (I 3053-N27 and GF17-33780L). SP and PS acknowledge support from the ERC through the AdG NLST. PS acknowledges support from the Royal Society through a University Research Fellowship (UHQT). RS and NB acknowledge financial support from the Swiss National Science Foundation (Starting grants DIAQ and QSIT). RS also acknowledges support from the SNSF project No. 200020\smallunderscore165843. AJS acknowledges support from an FQXi  ``Physics of what happens'' grant, via SVCF. 

\vspace{1cm} 

\noindent \emph{Note added:} After the completion of this work we have learned of an unpublished related earlier work by Caslav Brukner \cite{CaslavTalk} where the NBTS conditions \eqref{causality_one}, \eqref{eq:NBTSA} and \eqref{eq:NBTSB} were also formulated, and implications for the process matrix formalism were considered.

\bibliographystyle{unsrtnat}
\bibliography{processRefs_no_url.bib}

\newpage

\appendix

\section{Relation between classical equalities for indefinite relative timing} \label{ap:classical} 

In this appendix we focus on the case of indefinite relative timing and show that combining  the NBTS conditions with the relations 
\beq  p(a,b|x,y)+p(a,b|x',y')=p(a,b|x,y')+p(a,b|x',y)\label{eq:classicalequalityap} \eeq
when $a,b,x,y,x',y'$ are binary yields just one new equality. First note that the only non-trivial cases occur when $x \neq x'$ and $y\neq y'$. 
Consider the equality 
  \beq  p(0,0|0,0)+p(0,0|1,1)=p(0,0|0,1)+p(0,0|1,0) \eeq
by combining this with the NBTS conditions (Eqns.~\eqref{eq:NBTSA} and ~\eqref{eq:NBTSB}) we can show 
 \beq  p(0,1|0,0)+p(0,1|1,1)=p(0,1|0,1)+p(0,1|1,0) \eeq
as follows  
\begin{align}p(0,1|0,&0)+p(0,1|1,1) \nonumber \\
&\;=  p_A(0|y=0) +p_A(0|y=1) \nonumber\\
&\quad\qquad\qquad- p(0,0|0,0)-p(0,0|1,1)\nonumber\\
&\;=p_A(0|y=0) + p_A(0|y=1)\nonumber \\
&\quad\qquad\qquad- p(0,0|0,1)-p(0,0|1,0)\nonumber \\
&\;= p(0,1|1,0)+ p(0,1|0,1).
 \end{align} 
where we have used the fact that 
\begin{align} 
p_A(0|y=0) &= p(0,0|0,0) + p(0,1|0,0)  \nonumber \\
&= p(0,0|1,0) + p(0,1|1,0)
\end{align} 
Using a similar approach, we can prove the case with $a=1, b=0$, and by combining the NBTS conditions with one of these new equalities we can prove the $a=1, b=1$ case. 

In general if $a,b,x$ and $y$ can take $A,B, X$ and $Y$ different values respectively, then a similar argument shows that the number of new equalities given by \eqref{eq:classicalequalityap} is $(A-1)(B-1)(X-1)(Y-1)$.

\section{Characterisation of the classical polytope for indefinite relative timings}\label{ap:classical2}

In this appendix, in the case of indefinite relative timing, we show that when considering two parties and any number of inputs and outputs, the classical polytope is given by the intersection of the NBTS polytope with the the additional classicality conditions 
\begin{align}
\begin{split} \label{e:classicalequality}
p(a,b|x,y)+p&(a,b|x',y')\\
& =p(a,b|x,y')+p(a,b|x',y) ,
\end{split}
\end{align}
for all $a$, $b$, $x$, $y$, $x'$, $y'$. 

Let us consider that each party obtains one of $d$ outcomes, $a$, $b$ $\in \{0,\ldots,d-1\}$ and has $m$ inputs, $x$, $y$ $\in \{0,\ldots, m-1\}$. The classical polytope, in terms of vertices, is the convex hull of deterministic distributions, which fall into three families: the `actions overlap in time' family, comprised of vertices of the form
\begin{equation}\label{e:both const}
p(a,b|x,y) = \begin{cases} 1 & \text{if } a=\alpha,\quad b = \beta \\
0 & \text{otherwise} \end{cases}
\end{equation}
parametrised by $\alpha \in \{0,\ldots,d-1\}$ and $\beta \in \{0,\ldots,d -1\}$. Here $a = \alpha$ and $b = \beta$ are the deterministic outcomes, independent of $x$ and $y$; The `Alice's actions occur entirely before Bob's actions' family, comprised of vertices of the form
\begin{equation}\label{e:A const}
p(a,b|x,y) = \begin{cases} 1 & \text{if } a=\alpha,\quad b = \beta_x \\
0 & \text{otherwise} \end{cases}
\end{equation}
parametrised by $\alpha \in \{0,\ldots,d-1\}$ and $\beta_x \in \{0,\ldots,d -1\}$ for all $x$. Here $a = \alpha$, independent of $y$, and $b = \beta_x$; The `Bob's actions occur entirely before Alice's actions' family, comprised of vertices of the form
\begin{equation}\label{e:B const}
p(a,b|x,y) = \begin{cases} 1 & \text{if } a=\alpha_y,\quad b = \beta \\
0 & \text{otherwise} \end{cases}
\end{equation}
parametrised by $\alpha_y \in \{0,\ldots,d-1\}$ for all $y$, and $\beta \in \{0,\ldots,d -1\}$. Here $a = \alpha_y$, and $b = \beta$, independent of $x$.\footnote{We note that the `actions overlap in time' \eqref{e:both const} family is contained in both other families \eqref{e:A const} and \eqref{e:B const}, (in the case that $\alpha_y = \alpha$ or $\beta_x = \beta$). We present it as a separate sub-family for presentational purposes for what follows.}

In what follows, we will show that this polytope has an alternative characterisation, as the intersection of the NBTS polytope with the classicality conditions \eqref{e:classicalequality}. 

Note first that the classical polytope is contained in this intersection. This follows, since all of the above vertices satisfy the NBTS conditions (since they are a subset of the vertices of the NBTS polytope), and moreover can be seen to satisfy the classicality conditions \eqref{e:classicalequality}, due to the fact that in all three families at least one of the parties has a constant output. Finally any convex combination of the vertices also satisfy the same equalities.

What needs to be shown then is that any point $\PP = \{p(a,b|x,y)\}_{a,b,x,y}$ that satisfies the NBTS and  classicality conditions is contained inside the classical polytope. To do so, it suffices to show that any such $\PP$ can be written as a convex combination of the vertices of the classical polytope, i.e. of vertices of the form 
\eqref{e:both const} -- \eqref{e:B const}. In what follows, we will give an iterative procedure which  at every stage decomposes a point $\PP$ into a vertex of the classical polytope and a second point $\PP'$ that still satisfies the NBTS and classicality conditions. This procedure is shown to terminate, in which case an explicit decomposition is obtained.  

Consider a point $\PP = \{p(a,b|x,y)\}$ which satisfies the NBTS and classicality conditions. We start by identifying the smallest individual non-zero probability, i.e, the specific choice of outputs and inputs $a^*$, $b^*$, $x^*$, $y^*$ such that $p(a^*,b^*|x^*,y^*) \leq p(a,b|x,y)$ for all $a,b,x,y$ such that $p(a,b|x,y) \neq 0$. Let us denote $\epsilon = p(a^*,b^*|x^*,y^*)$. 

We next check whether, for any value $x \neq x^*$
\begin{equation}\label{e:x 0}
p(a^*,b^*|x,y^*) = 0,
\end{equation}
or whether for any value $y \neq y^*$
\begin{equation}\label{e:y 0}
p(a^*,b^*|x^*,y) = 0.
\end{equation}
There are four possibilities: (i) there is no $x$ such that \eqref{e:x 0} is satisfied and no $y$ such that \eqref{e:y 0} is satisfied; (ii) there is no $x$ such that \eqref{e:x 0} is satisfied but a non-empty subset of $y$ such that \eqref{e:y 0} is satisfied; (iii) there is no $y$ such that \eqref{e:y 0} is satisfied but a non-empty subset of $x$ such that \eqref{e:x 0} is satisfied; (iv) there is simultaneously a non-empty subset of $x$ such that \eqref{e:x 0} is satisfied and a non-empty subset of $y$ such that \eqref{e:y 0} is satisfied.  

Note that the last possibility is in fact impossible. It would imply in particular that there is an $x'$ and $y'$ such that $p(a^*,b^*|x',y^*) = 0$ and $p(a^*,b^*|x^*,y') = 0$. From the classicality conditions \eqref{e:classicalequality} applied to $x'$ and $y'$ in conjunction with $x^*$ and $y^*$, it would then follow that $p(a^*,b^*|x^*,y^*) = 0$, since
\begin{align}
p(a^*,b^*|x^*,y^*) &+ p(a^*,b^*|x',y') \\
&\;= p(a^*,b^*|x',y^*) + p(a^*,b^*|x^*,y')\nonumber \\
&\; = 0,
\end{align}
but by assumption $p(a^*,b^*|x^*,y^*) = \epsilon > 0$, which is a contradiction.

Let us assume first then that case (i) holds, i.e.~that $p(a^*,b^*|x,y^*) \geq \epsilon$ for all $x$ and $p(a^*,b^*|x^*,y) \geq \epsilon$ for all $y$.\footnote{Note that, since $p(a^*,b^*|x^*,y^*) = \epsilon$ was assumed to be the smallest non-zero probability, if $p(a^*,b^*|x,y^*) \neq 0$ then necessarily $p(a^*,b^*|x,y^*) \geq \epsilon$, and similarly for $p(a^*,b^*|x^*,y)$.} It then follows that
\begin{equation}\label{e:geq eps}
p(a^*,b^*|x,y) \geq \epsilon
\end{equation}
for all $x$, $y$. Indeed, let us assume that this were not the case, i.e.~that for some choice $x'$ and $y'$, $p(a^*,b^*|x',y')~=~0$. From the classicality conditions \eqref{e:classicalequality} it would then follow that
\begin{align}
p(a^*,b^*|x^*,y') &+ p(a^*,b^*|x',y^*) \nonumber\\
&\;= p(a^*,b^*|x^*,y^*) + p(a^*,b^*|x',y')  \nonumber\\
 &\;= \epsilon.
\end{align} 
However this is impossible, since both $p(a^*,b^*|x^*,y')$ and $p(a^*,b^*|x',y^*)$ are by assumption non-vanishing and at least as large as $\epsilon$. Thus in case (i), we see that $p(a^*,b^*|x,y) \geq \epsilon$ for all $x$, $y$. 

Consider now the point $\PP_c = \{p_c(a,b|x,y)\}$ that is a vertex of the classical polytope from the family \eqref{e:both const} with $\alpha = a^*$, $\beta = b^*$, i.e. the deterministic distribution where $a = a^*$, $b = b^*$ and $p_c(a,b|x,y) = \delta_{a,a^*}\delta_{b,b^*}$. 

The above shows that the distribution $\PP$ can be written as
\begin{equation}
\PP = \epsilon \PP_c + (1-\epsilon) \PP'
\end{equation}
if $\epsilon < 1$, or $\PP = \PP_c$ if $\epsilon = 1$, where $\PP' = \{p'(a,b|x,y)\}$ is some other point. Indeed, the above analysis guarantees that $p'(a,b|x,y) \geq 0$ for all $a$, $b$, $x$, $y$, since $p'(a^*,b^*|x,y) = (p(a^*,b^*|x,y) - \epsilon)/(1-\epsilon) \geq 0$ and  $p'(a,b|x,y) = p(a,b|x,y)/(1-\epsilon) \geq 0$ if $a \neq a^*$ or $b \neq b^*$. Moreover it also satisfies the NBTS conditions, normalisation of probabilities, and the classicality conditions \eqref{e:classicalequality}, due to linearity. Finally, it has the important property that $p'(a^*,b^*|x^*,y^*) = 0$, i.e. $\PP'$ has at least one more vanishing probability than $P$.  

Thus, when case (i) occurs, either $\PP$ was an `actions overlap in time' vertex of the  classical polytope (when $\epsilon = 1)$, or it is possible to write it as a convex combination of such a vertex and a second distribution. In the former case, we have achieved the goal of showing that $\PP$ is contained in the classical polytope, while in the second case we can now re-start the above procedure, focusing on $\PP'$ instead of $\PP$. 

Let us now assume that case (i) does not hold, but rather case (ii), i.e. $p(a^*,b^*|x,y^*) \geq \epsilon$ for all $x$ and $p(a^*,b^*|x^*,y) = 0$ for some non-empty subset of $y$. 

We will show that for each value of $y$, there is an outcome $a = a_y$ such that $p(a_y,b^*|x,y) \geq \epsilon$ for all $x$.

%Consider first those  $y$ such that $p(a^*,b^*|x^*,y) \geq \epsilon$. Then since $p(a^*,b^*|x^*,y^*) = \epsilon$ and $p(a^*,b^*|x,y^*) \geq \epsilon$ for all $x$ by assumption, it follows that $p(a^*,b^*|x,y) \geq \epsilon$, since from the classicality condition 
%\begin{align}
%p(a^*,b^*|x,y) &= p(a^*,b^*|x^*,y) + p(a^*,b^*|x,y^*) \nonumber \\&\quad - %p(a^*,b^*|x^*,y^*), \nonumber \\
%&\geq \epsilon
%\end{align}
%For these $y$ we can therefore take $a_{y} = a^*$ .

%Consider now the subset of $y$ such that $p(a^*,b^*|x^*,y) = 0$. 
The only way that it would be impossible to find, for some $y'$, an $a_{y'}$ such that $p(a_{y'},b^*|x,y') \geq \epsilon$ for all $x$, would be if for each value of $a$, there was an input $x_a$ such that $p(a,b^*|x_a,y') = 0$.  Indeed, in this case, there is no suitable choice for $a_{y'}$, since every choice is ruled out by the input $x_a$. We will now show that this cannot occur. 

Assuming that the above can happen, that an $x_a$ exists for each $a$ such that $p(a,b^*|x_a,y') = 0$, then from the classicality conditions \eqref{e:classicalequality} it would follow that for all $a$,
\begin{align}\label{e:imp 1}
p(a,b^*|x^*,y^*) = p(a,b^*|x_a,y^*) &+ p(a,b^*|x^*,y').
\end{align} 
From the NBTS condition of Alice, it holds that
\begin{equation}
\sum_a p(a,b^*|x^*,y^*) = \sum_a p(a,b^*|x^*,y'). 
\end{equation}
Substituting \eqref{e:imp 1} into this NBTS condition, this would therefore imply that
\begin{equation}
\sum_a p(a,b^*|x_a,y^*) = 0
\end{equation}
which in turn would imply that $p(a,b^*|x_a,y^*) = 0$ for all $a$. This however cannot occur, since for the choice $a = a^*$ it would imply that $p(a^*,b^*|x_{a^*},y^*) = 0$, however by assumption of case (ii), $p(a^*,b^*|x,y^*) \geq \epsilon$ for all $x$. This contradiction shows that no such $x_a$ can exist.

In summary, for each value of $y$ there is an outcome $a = a_y$ such that $p(a_y,b^*|x,y) \geq \epsilon$ for all $x$. 

Thus, similarly to case (i), if we consider now the point $\PP_c$ which is a classical vertex from the family \eqref{e:B const} with $\alpha_y = a_y$ and $\beta = b^*$, i.e. such that $p_c(a,b|x,y) = \delta_{a,a_y}\delta_{b,b^*}$, then the above implies that it is possible to decompose $\PP$ as
\begin{equation}
\PP = \epsilon \PP_c + (1-\epsilon) \PP'
\end{equation}
if $\epsilon < 1$, or $\PP = \PP_c$ if $\epsilon = 1$, where again $\PP'$ is some other (positive) distribution which, as well as still satisfying the NBTS, normalisation and classicality conditions, has the analogous property to before that $p'(a^*,b^*|x^*,y^*) = 0$, i.e. has at least one more vanishing probability than $p(a,b|x,y)$. 

Thus, when case (ii) occurs, $\PP$ is shown either to be equal to an `Alice's actions occur entirely before Bob's actions' vertex of the classical polytope, or it is possible to write it as a convex combination of a such a vertex and a second distribution. Once again, in the former case we have the desired decomposition, and in the latter we can re-start the above procedure, focusing on $\PP'$ instead of $\PP$. 

Finally, case (iii) is identical to case (ii), except the role of Alice and Bob is reversed. That is, if we are in case (iii), for each value of $x$ it is always possible to find an outcome $b = b_x$ such that $p(a^*,b_x|x,y) \geq \epsilon$ for all $y$. Subsequently $\PP$ can be decomposed as $\PP = \epsilon \PP_c + (1-\epsilon) \PP'$ if $\epsilon < 1$ or $\PP = \PP_c$ if $\epsilon = 1$, where now $\PP_c$ is a vertex from the family \eqref{e:A const} with $\alpha = a^*$ and $\beta_x = b_x$, and $\PP'$ has at least one more vanishing probability compared to $\PP$. 

In conclusion, given any distribution $\PP = \{p(a,b|x,y)\}_{a,b,x,y}$ that satisfies the NBTS, normalisation and classicality conditions, by iterating the above procedure, a sequence of decompositions are generated, 
%\begin{widetext}
\begin{align}
\PP &= \epsilon^{(1)}\PP_c^{(1)}+ (1-\epsilon^{(1)})\PP^{(1)} \nonumber \\
&= \epsilon^{(1)}\PP_c^{(1)} + (1-\epsilon^{(1)})[\epsilon^{(2)}\PP_c^{(2)} + (1-\epsilon^{(2)})\PP^{(2)}] \nonumber \\
&\hspace{5cm}\vdots \nonumber \\
&= \epsilon^{(1)}\PP_c^{(1)} + (1-\epsilon^{(1)})\epsilon^{(2)}\PP_c^{(2)} + \ldots + \prod_{i=1}^{k}(1-\epsilon^{(i)})\PP^{(k)} \nonumber \\
&\hspace{5cm}\vdots
\end{align}
%\end{widetext}
such that the `remainder' $\PP^{(k)}$ is a valid distribution (positive, normalised and satisfying the NBTS conditions) and has at least one more vanishing probability than the previous remainder $\PP^{(k-1)}$. Since there are only at most $d^2m^2$ non-vanishing probabilities, the sequence cannot carry on indefinitely and it must be the case that after a finite number $N$ of iterations the procedure terminates. This happens exactly when the smallest probability of the remainder is in fact unity, $\epsilon^{(N)} = 1$, in which case, from the above analysis, we are guaranteed that it will be a vertex of the classical polytope, $\PP^{(N)} = \PP^{(N)}_c$.  When the procedure terminates at this stage we have thus obtained an explicit convex decomposition of the original distribution $\PP$ into vertices of the classical polytope,
\begin{equation}
\PP = \epsilon^{(1)}\PP_c^{(1)} + (1-\epsilon^{(1)})\epsilon^{(2)}\PP_c^{(2)} + \ldots + \prod_{i=1}^{N-1}(1-\epsilon^{(i)})\PP_c^{(N)} \nonumber \\
\end{equation}

As such, any $\PP$ satisfying the NBTS, normalisation, and classicality conditions is contained inside the classical polytope. The classical polytope for two parties is thus characterised as the intersection of the NBTS polytope with the classicality conditions \eqref{e:classicalequality}. An alternative way of saying this, is that the NBTS, normalisation and classicality conditions are necessary and sufficient conditions for a point $\PP$ to be classical.

\section{Two-time quantum states} \label{ap:two-time} 

In this appendix we consider which  two-time quantum states obey NBTS, in the sense that they give rise to probabilities obeying the NBTS constraints.  In particular, we first consider a single party (Alice) who has a lab into which a quantum state enters, they perform a quantum measurement on that state (obtaining result $a$), then apply a quantum channel to the state (labelled by $x$), before sending it out of their laboratory. At the entrance and exit of their laboratory, a general two-time quantum state $\eta$  is prepared.  If $p(a|x) =p(a)$ for all choices of measurement and channel by Alice, then we say that the two-time state $\eta$ obeys NBTS. 

We then consider two parties, Alice and Bob, each of whom has a laboratory into which a quantum state enters, they perform a measurement on it (with outputs $a$ and $b$ respectively), then apply a channel to the state (labelled by $x$ and $y$) and then output the resultant system from their laboratory. Outside their laboratories, Alice and Bob's inputs and outputs are prepared in an arbitrary two-time state. If $p(a,b|x,y)$ satisfies the NBTS conditions for a given scenario, then we say that the two-party state obeys NBTS. 

We prove the following results: 
(i) for a single party, the only  states which obey NBTS correspond to standard (pre-selected) quantum states with no post-selection, or with trivial post-selection (i.e. post-selection on an ancilla for which the probability of success is independent of Alice's operations).  
For two parties: (ii) the  states which obey the NBTS conditions for indefinite relative timing correspond to the \emph{linear two time states} defined in \cite{Silva2017}, which are equivalent to process matrices \cite{Oreshkov2012}. If these states are simulated via experimental post-selection, then the probability of success in the post-selection is independent of each party's operations. (iii) Linear two-time states satisfy the equality \eqref{eq:classicalequality} satisfied by classical correlations. (iv) the states which obey the NBTS conditions with definite and parallel relative timing correspond to standard pre-selected quantum states with no post-selection, or with trivial post-selection. (v) the states which obey the NBTS conditions with definite relative timing and Alice before Bob correspond to states which do not involve a post-selection for Bob (or a trivial post-selection). (vi) In all of the above cases, the states can only produce classical correlations.

\subsection{Review of pre- and post-selected formalism} 

Here we briefly review the formalism for pre- and post- selected quantum states presented in \cite{Ralph2014, Silva2017}. 

For an arbitrary process (which may be a state, measurement or channel), we associate a Hilbert space $ \mathcal{H}^{\mathcal{O}} \otimes \mathcal{H}_{\mathcal{O}^{\dagger}}$ to every output $O$, where $\mathcal{H}^{\mathcal{O}}$ (with raised index) is a standard vector space (represented by a ket) and  $\mathcal{H}_{\mathcal{O}^{\dagger}}$ (with lowered index)  is a dual vector space (represented by a bra). Similarly, we associate a Hilbert space $ \mathcal{H}_{\mathcal{I}} \otimes \mathcal{H}^{\mathcal{I}^{\dagger}}$ to every input $I$. 

The mathematical object associated with that process is then a vector in the tensor product space of all of the output and input spaces. Composition of two processes is given by the $\bullet$ operation, which connects vectors and dual vectors with the same label to give a scalar (i.e. $_{\mathcal{A}}\bra{\psi} \bullet \ket{\phi}^{\mathcal{A}} = \inner{\psi}{\phi}$) and performs the tensor product on  vectors or dual vectors with different labels. All allowed physical processes correspond to `positive' vectors in the appropriate Hilbert space. In particular given a process $C_{A_1}^{A_2}$ with input $A_1$ and output $A_2$, then $(v \otimes v^{\dagger}) \bullet C_{A_1}^{A_2} \geq 0$ for all $v \in \mathcal{H}^{\mathcal{A}_1} \otimes \mathcal{H}_{\mathcal{A}_2}$ (with Hermitian conjugate $v^{\dagger} \in \mathcal{H}_{\mathcal{A}_1^{\dagger}} \otimes \mathcal{H}^{\mathcal{A}_2^{\dagger}}$). 

Positivity is the only condition for a two-time state to be physically achievable. Channels (implemented without post selection) must also satisfy an additional condition, which corresponds to them being trace-preserving. For a channel  $C_{A_1}^{A_2}$  from $A_1$ to $A_2$ we require that $\ident_{A_2} \bullet  C_{A_1}^{A_2} = \ident_{A_1}$, where $\ident_{A_1} = \sum_i \ket{i}^{\mathcal{A}_1^{\dagger}} \otimes \bra{i}_{\mathcal{A}_1}$ (i.e. they must be future-identity preserving). A measurement is described by a set of processes $M_a$ corresponding to the output $a$, such that $M= \sum_a M_a$ is a valid channel. For the sake of familiarity, it is sometimes helpful to move between vectors in the pre- and post-selected formalism (such as $\rho^{A} = \sum_{ij} \rho_{ij} \ket{i}^{\mathcal{A}} \otimes \bra{j}_{\mathcal{A}^{\dagger}}$) and the corresponding operator in the standard quantum formalism, which we will denote with a `hat' (e.g. $\hat{\rho}^{A} = \sum_{ij} \rho_{ij} \ket{i} \bra{j}$).

In any pre- and post-selected scenario, the joint probability of obtaining any particular set of measurement outcomes can be obtained by composing all of the processes (with the measurements having those particular outcomes), and dividing by the same quantity summed over all outcomes (which corresponds with replacing the individual measurement outcomes with the corresponding measurement channels). For example,  given a two-time state state $\eta^{A_1}_{A_2}$ and a measurement $(M_a)_{A_1}^{A_2}$, the probability of obtaining outcome $a$ is given by 
\begin{equation} 
p(a)= \frac{ \eta \bullet M_a}{\eta \bullet M}
\end{equation} 
where $M= \sum_a M_a$. Note that the overall normalisation of $\eta$ is not physically relevant.  If $\eta$ is simulated via experimental post-selection, then the denominator of this expression ($\eta \bullet M$) is proportional to the probability of the post-selection succeeding (with respect to changes in $M$). 

\subsection{No-backwards-in-time signalling states} 

Within the framework described in the previous section, we now define the class of two-time states obeying NBTS. Note that in this section we consider that Alice's transformation depends only upon $x$ and not on $a$ for simplicity (and similarly for Bob). However, the measurements performed  could append an ancilla to the system containing the measurement result, which could then be conditioned on in the transformation. In this way, any dependence of the transformation on the measurement results is implicitly included, and hence there is no loss of generality. 

\begin{definition}  \textbf{No-backwards-in-time-signalling for  two-time states}. We say that a two-time state $\eta$ for a single party obeys  NBTS if the outcome probabilities $p(a|x)$ are independent of $x$  when we perform any (non-destructive, trace-preserving) measurement $J_a$ followed by a trace-preserving channel chosen from a set $L_x$. In particular we demand
\begin{align} \label{eq:probs}
\qquad p(a|x) &\equiv \frac{L_x \bullet J_a \bullet \eta }{ L_x \bullet J \bullet \eta} =p(a)
\end{align} 
i.e. it is independent of $x$, 
where $\eta= \eta^{A_1}_{A_3 }$, $J_a= (J_a)^{A_2 }_{A_1 }$, with $J=\sum_a J_a$, and $L_x=(L_x)^{A_3 }_{A_2}$.  

\medskip

When considering a pre- and post- selected state $\eta$ of two parties, in the case of indefinite relative timing, we say that  it obeys NBTS if 
\begin{align}
p(a|x, y) &\equiv \frac{L_x \bullet J_a \bullet L'_y \bullet J'  
		\bullet \eta }{ L_x \bullet J \bullet L'_y \bullet J'  \bullet \eta} =p(a|y) \\ 
 p(b|x, y) &\equiv \frac{L_x \bullet J \bullet L'_y \bullet J'_b  
 		\bullet \eta }{ L_x \bullet J \bullet L'_y \bullet J'  \bullet \eta} =p(b|x) 
\end{align} 
where $J_a$ and $L_x$ are defined as above, $\eta= \eta^{A_1  B_1 }_{A_3  B_3 }$, $J'_b= (J'_b)^{B_2}_{B_1}$ with $J'=\sum_b J_b$, and $L'_y=(L_y)^{B_3 }_{B_2 }$.  \\

For the case of definite parallel timing, the same state $\eta$ obeys NBTS if 
\begin{align}
p(a|x, y) &\equiv \frac{L_x \bullet J_a \bullet L'_y \bullet J'  
		\bullet \eta }{ L_x \bullet J \bullet L'_y \bullet J'  \bullet \eta} =p(a) \\ 
 p(b|x, y) &\equiv \frac{L_x \bullet J \bullet L'_y \bullet J'_b  
 		\bullet \eta }{ L_x \bullet J \bullet L'_y \bullet J'  \bullet \eta} =p(b) \,.
\end{align} 

Finally, for the case of definite relative timing ($A\rightarrow B$), the two-time state $\eta$ obeys NBTS if 
\begin{align}
p(a|x, y) &\equiv \frac{L_x \bullet J_a \bullet L'_y \bullet J'  
		\bullet \eta }{ L_x \bullet J \bullet L'_y \bullet J'  \bullet \eta} =p(a) \\ 
 p(b|x, y) &\equiv \frac{L_x \bullet J \bullet L'_y \bullet J'_b  
 		\bullet \eta }{ L_x \bullet J \bullet L'_y \bullet J'  \bullet \eta} =p(b|x) \,.
\end{align}  
\end{definition} 

\begin{remark} \label{rmk:1}
Note that  if $\eta$ is a state for two parties which obeys NBTS, then the marginal state
\begin{equation} 
\eta\,^{\!(A)} = L_y' \bullet J' \bullet \eta
\end{equation} 
for a single party obeys NBTS for every $y$. 
\end{remark} 

We now prove the first result, concerning single-party states which obey NBTS. 

\medskip

\begin{theorem} \label{thm:1}  \textbf{ States for a single party satisfying no-backwards-in-time-signalling correspond to states without post-selection, or with trivial post-selection.} If the two-time state $\eta= \mathlarger{\eta}^{A_1 }_{A_3 }$ for a single party obeys no-backwards-in-time-signalling, then
\begin{equation} 
\eta^{A_1 }_{A_3 }= \rho^{A_1 } \otimes \ident_{A_3},
\end{equation} 
where $\rho$ is a positive vector.
\end{theorem} 
\proof 
Any $\eta$ can be decomposed as 
\begin{equation} \label{eq:eta} 
\eta = \sum_{i, j} \ket{i}^{a_1} \otimes \bra{j}_{a_1^{\dagger}} \otimes (B_{ij})_{A_3}.
\end{equation} 
Consider a  measurement $J_a$ followed by a set of channels $L_x$, and two outcomes $l$ and $k$.
We now consider three particular choices of measurements 
\begin{enumerate}[label=(\roman*)]
\item \label{case:1}
$J_a$ corresponds to a measurement in the computational basis, followed by the preparation of the $\ket{0}$ state:
\begin{equation} 
J_a = \ket{0}^{a_2} \otimes \bra{a}_{a_1} \otimes \ket{a}^{a_1^{\dagger}} \otimes \bra{0}_{a_2^{\dagger}}.
\end{equation} 
$L_x$ corresponds to a unitary channel $U_x$ with $x$ labelling all possible unitaries\footnote{or alternatively a sufficient set of $x$ such that  $U_x\proj{0} U_x^{\dagger}$ form an operator basis}, 
\begin{equation} 
L_x = (U_x)^{a_3}_{a_2} \otimes (U_x^{\dagger})^{a_2^{\dagger}}_{a_3^{\dagger}}.
\end{equation} 
In this case  
\begin{align} \label{eq:comparison}
 L_x \bullet J_a \bullet \eta &= \sum_{i,j} \inner{a}{i} \, \inner{j}{a} \,\bra{0} \hat{U}_x^{\dagger} \hat{B}_{ij} \hat{U}_x \ket{0} \nonumber \\ &= \bra{0}\hat{U}_x^{\dagger} \hat{B}_{aa} \hat{U}_x \ket{0} 
\end{align} 
from (\ref{eq:probs}) we therefore obtain
\begin{equation} 
 \frac{\bra{0} \hat{U}_x^{\dagger} \hat{B}_{aa} \hat{U}_x \ket{0} }{ \sum_k \bra{0} \hat{U}_x^{\dagger} \hat{B}_{kk} \hat{U}_x \ket{0}} =p(a)
\end{equation} 
and hence, defining $B= \sum_k  B_{kk}$, 
\begin{equation} 
\bra{0} \hat{U}_x^{\dagger} (\hat{B}_{aa} - p(a)\hat{B})  \hat{U}_x \ket{0}  = 0 
\end{equation} 
As this holds for all $\hat{U}_x$ we find  
\begin{equation} \label{eq:case1} 
B_{aa} =  p(a)B. 
\end{equation} 

\item \label{case:2}
The second measurement we consider is similar to the first, but two of the measurement outcomes correspond to the states  $\ket{\pm} = \frac{1}{\sqrt{2}} \left(\ket{r} \pm \ket{s}\right)$ for arbitrary $r$ and $s$. The other  measurement elements can be taken to be in the computational basis. 
\begin{align} 
J_\pm &= \ket{0}^{a_2} \otimes \bra{\pm}_{a_1} \otimes \ket{\pm}^{a_1^{\dagger}} \otimes \bra{0}_{a_2^{\dagger}},\nonumber \\  J_{a \neq r,s} &= \ket{0}^{a_2} \otimes \bra{a}_{a_1} \otimes \ket{a}^{a_1^{\dagger}} \otimes \bra{0}_{a_2^{\dagger}}.
\end{align}
$L_x$ is the same as in case 1. Proceeding as before we obtain 
\begin{equation}   \label{eq:case2} 
\frac{1}{2} \left( B_{rr} + B_{rs} + B_{sr} +  B_{ss}  \right) =  p(+) B. 
\end{equation} 

\item \label{case:3}
The third measurement we consider is the same as case \ref{case:2}, but two of the measurement outcomes correspond to $\ket{\pm i} = \frac{1}{\sqrt{2}} \left(\ket{r} \pm i \ket{s}\right)$, with 
\begin{align} 
J_{\pm i} &= \ket{0}^{a_2} \otimes \bra{\pm i}_{a_1} \otimes \ket{\pm i}^{a_1^{\dagger}} \otimes \bra{0}_{a_2^{\dagger}}, \nonumber \\ J_{a \neq r,s} &= \ket{0}^{a_2} \otimes \bra{a}_{a_1} \otimes \ket{a}^{a_1^{\dagger}} \otimes \bra{0}_{a_2^{\dagger}}.
\end{align}
$L_x$ is the same as in case 1. Proceeding as before we obtain 
\begin{equation}   \label{eq:case3} 
\frac{1}{2} \left( B_{rr} +i  B_{rs} -i B_{sr} +  B_{ss}  \right) =  p(+i) B. 
\end{equation} 

\end{enumerate} 
Combining equations (\ref{eq:case1}), (\ref{eq:case2}) and (\ref{eq:case3}) for all $a, r, s$ it is straightforward to see that all $B_{ij}$ are proportional to $B$. Writing  $B_{ij}=c_{ij} B$ it follows from (\ref{eq:eta}) that $\eta$ has the product form  
\begin{equation} 
\eta = C^{A_1} \otimes  B_{A_3} 
\end{equation} 
where 
\begin{equation} 
C^{A_1}= \sum_{i,j} c_{ij}  \ket{i}^{a_1} \otimes \bra{j}_{a_1^{\dagger}}.
\end{equation} 
We now show that $B \propto I$. Consider performing the measurement given by 
\begin{equation} 
J_a = \frac{1}{d_{a_2}} \ket{a}^{a_2} \otimes \bra{a}_{a_2^{\dagger}} \otimes \ident_{A_1}.
\end{equation} 
This measurement corresponds to throwing away the input state, outputting a random number $a$ from 1 to $d_{A_2}$ and then outputting the pure state $\ket{a}_{a_2}$. We then perform the  channel $ L_x$ as above. This gives
\begin{equation} 
 L_x \bullet J_a \bullet \eta =  \frac{1}{d_{A2}}\tr(\hat{C}) \bra{a} \hat{U}_x^{\dagger} \hat{B} \hat{U}_x \ket{a} 
\end{equation} 
and hence, 
\begin{equation} 
p(a)= \frac{\tr(\hat{C}) \bra{a} \hat{U}_x^{\dagger} \hat{B} \hat{U}_x \ket{a} }{\tr(\hat{C})  \tr(\hat{B}) }. 
\end{equation} 
As this holds for all $U_x$ it follows that $B= \lambda \ident$ for some constant $\lambda$. Hence 
\begin{equation} 
\eta^{A_1 }_{A_3 }= \rho^{A_1 } \otimes \ident_{A_3}
\end{equation} 
as desired, where $\rho=\lambda C$. 
\proofend

\bigskip 

We now use this result to prove that for two parties, the two-time states which obey NBTS are equivalent to the linear two time states (which were previously shown to be equivalent to process matrices in \cite{Silva2017}). 

\begin{theorem} \label{thm:constantdenom} \textbf{ For the case of indefinite relative timing, a two-party, two-time state obeys NBTS if and only if it is  proportional to (and thus physically equivalent to) a linear two-time state.} Given any two  trace-preserving measurements $ (J_a)^{A_3}_{A_1 }$ and  $ (K_b)^{B_3 }_{B_1 }$, with $J=\sum_a J_a$, and $K=\sum_b K_b$, a two-time state $\eta =\mathlarger{\eta}^{\mathsmaller{A_1 B_1}}_{\mathsmaller{A_3 B_3} }$ is linear if 
\begin{equation} 
p(a, b)= J_a \bullet K_b \bullet \mathlarger{\eta} 
\end{equation} 
\end{theorem} 
\proof We first show that any causal two-time state is proportional to a linear two-time state. Consider a channel for each party corresponding to doing nothing to the state. These are given by 
\begin{equation} 
L_0 = \mathlarger{\ident}
_{a_2}^{a_3} \otimes \ident_{a_3^{\dagger}}^{a_2^{\dagger}} \qquad M_0 = 
\mathlarger{\ident}
_{b_2}^{b_3} \otimes 
	\mathlarger{\ident}
	_{b_3^{\dagger}}^{b_2^{\dagger}}.
\end{equation}  
Relabelling the spaces on which the measurements act to get  $J_a' = (J_a)_{A_1}^{A_2}$ and $K'_b = (K_b)^{B_2 }_{B_1 }$, we note that 
\begin{equation} 
J \bullet K \bullet \eta =  L_0 \bullet J'   \bullet M_0\bullet K' \bullet \eta'.
\end{equation}  
If $\eta'$ is a two-party causal state, it follows from Remark \ref{rmk:1} and Theorem \ref{thm:1} that the marginal state satisfies 
\begin{equation}  
\eta\,^{\!(A)} = M_0 \bullet K'  \bullet \eta =\rho^{A_1 } \otimes \ident_{A_3}
\end{equation}
and thus
\begin{align} 
J \bullet K \bullet \eta &=L_0 \bullet J' \bullet \left( \rho \otimes \ident_{A_3} \right) \\
& = J' \bullet \ident_{A_2} \bullet \rho \\
&= \ident_{A_1} \bullet \rho
\end{align} 
where we have used the fact that $J'$ and $L_0$ are trace-preserving channels. Hence $J \bullet K \bullet \eta $  is independent of $J$. Following the same argument with the parties swapped shows that $J \bullet K \bullet \eta$ is also independent of $K$ and is thus a constant $c$. Now the state $\mathlarger{\eta}_{\mathsmaller{W}}= \eta/c$ satisfies $J \bullet K \bullet \mathlarger{\eta}_{\mathsmaller{W}}$=1 and thus
\begin{equation} 
p(a, b)=J_a \bullet K_b \bullet \mathlarger{\eta}_{\mathsmaller{W}}
\end{equation} 
and hence $\mathlarger{\eta}_{\mathsmaller{W}}$ is a linear two-time state. To prove the converse we use the Theorem given in \cite{Silva2017}, which implies\footnote{To obtain the first expression, consider the first equation in (30) from \cite{Silva2017}. Then set $C=L_x$ and $\tilde{C} = L_{x'}$,  replace $K$ by $M_y \bullet K$, and take $\bullet J_a$ on both sides. The second equation can be obtained by a similar argument with the parties swapped.}  that for a linear two-time state $\mathlarger{\eta}_{\mathsmaller{W}}$ 
\begin{align}
L_x \bullet J_a \otimes M_y \bullet K \bullet \mathlarger{\eta}_{\mathsmaller{W}}&= L_{x'} \bullet J_a \otimes M_y \bullet K \bullet 
\mathlarger{\eta}_{\mathsmaller{W} } \\
L_x \bullet J \otimes M_y \bullet K_b \bullet \mathlarger{\eta}_{\mathsmaller{W}}&= L_x  \bullet J \otimes M_{y'} \bullet K_b \bullet 
\mathlarger{\eta}_{\mathsmaller{W}}
\end{align} 
where $J_a$ corresponds to Alice's measurement and $K_b$ to Bob's measurement, and $L_x, M_y, J=\sum_a J_a$ and $K = \sum_b K_b$ correspond to completely positive trace preserving maps. These are just a representation of the NBTS conditions equivalent to 
\begin{align} 
\sum_b p(a ,b |x ,y) = \sum_b p(a, b |x' ,y) \\
\sum_a p(a ,b |x ,y) = \sum_a p(a ,b |x ,y'). 
\end{align} 
\proofend

Finally, the third equation of (30) in \cite{Silva2017} implies that for a linear two-time state 
\begin{equation} 
(L_x - L_{x'}) \bullet J_a \bullet (M_y - M_y') \bullet K_b \bullet \mathlarger{\eta}_{\mathsmaller{W}} = 0 
\end{equation} 
which correspond to the equalities obeyed by the classical polytope \eqref {eq:classicalequalityap}  
\beq  p(a,b|x,y)+p(a,b|x',y')=p(a,b|x,y')+p(a,b|x',y). \eeq

Now, moving on to the case of definite relative timings, we prove the following: 

\begin{theorem} \label{thm:constantdenom} \textbf{ A two-party, two-time state obeys the NBTS conditions with a fixed relative time ordering where both experiments occur in parallel if and only if it is  proportional to (and thus physically equivalent to) a pre-selected only state.} If the two-time state $\eta= \mathlarger{\eta}^{A_1 B_1}_{A_3 B_3}$ obeys no-backwards-in-time-signalling in this setting, then
\begin{equation} 
\eta^{A_1 B_1}_{A_3 B_3}= \rho^{A_1 B_1} \otimes \ident_{A_3}\otimes \ident_{B_3},
\end{equation} 
where $\rho$ is a positive vector.
\end{theorem} 
\proof The NBTS conditions in this scenario, at the level of two-time states are
\begin{subequations}
\begin{align}
(L \bullet J_a \otimes M )\bullet \eta &= (L' \bullet J_a \otimes M) \bullet 
\eta \label{e:a}\\
(L \bullet J_a \otimes M )\bullet \eta &= (L \bullet J_a \otimes M') \bullet 
\eta \label{e:b}\\
(L \otimes M \bullet K_b )\bullet \eta &= (L \otimes M' \bullet K_b) \bullet \eta \label{e:c}\\
(L \otimes M \bullet K_b ) \bullet \eta &= (L' \otimes M \bullet K_b) \bullet \eta \label{e:d}
\end{align} 
\end{subequations}
where $ J_a$ and $ K_b$ are arbitrary trace-preserving measurements, and  $ L$, $ L'$, $ M$ and $M'$ are arbitrary trace-preserving channels. 

Note first that $L \bullet J_a$ is an arbitrary positive vector.\footnote{The only requirement on $L \bullet J_a$ is that $L \bullet J_a \bullet \eta \geq 0$ for all $\eta$, which can be seen as a `positivity' requirement.} The only way that \eqref{e:b} can be true for an arbitrary positive vector $L\bullet J_a$ is if it is true at the level of the two-time state itself\footnote{In particular, it is always possible to construct a basis of positive vectors, and the only way this equation can hold true for a basis is if it holds true in general}, i.e. if
\begin{equation}\label{e:new condition}
M \bullet \eta =  M' \bullet \eta
\end{equation}
where we recall that $M = M^{B_3}_{B_1}$ and $M' = (M')^{B_3}_{B_1}$. 

\noindent Now, let us consider the specific choice 
\begin{align}
M^{B_3}_{B_1} &= T^{B_3}_{B_1} \nonumber \\
(M')^{B_3}_{B_1} &= T^{B_3}_{B_1} + \epsilon(T^{B_3}_{B_2}-  I^{B_3}_{B_2})\bullet X^{B_2}_{B_1}
\end{align}
where $T$ is the throw-away-and-replace channel defined as
\begin{align}
T^{B_3}_{B_1} = \frac{1}{d_{b_3}} \ident^{B_3} \bullet \ident_{B_1}
\,,
\end{align}
$I$ is the identity channel; $X$ is an arbitrary vector\footnote{this vector must be `Hermitian' in the sense that it produces real numbers when acting on valid two-time states $\eta^{\mathsmaller{B_1}}_{\mathsmaller{B_2}}\bullet X_{B_1}^{B_2}\;\in \mathbb{R}\,,\,\forall \eta^{\mathsmaller{B_1}}_{\mathsmaller{B_2}}\,.$
%
%
% All allowed physical processes correspond to `positive' vectors in the appropriate Hilbert space. In particular given a process $C_{A_1}^{A_2}$ with input $A_1$ and output $A_2$, then $(v \otimes v^{\dagger}) \bullet C_{A_1}^{A_2} \geq 0$ for all $v \in \mathcal{H}^{\mathcal{A}_1} \otimes \mathcal{H}_{\mathcal{A}_2}$ (with Hermitian conjugate $v^{\dagger} \in \mathcal{H}_{\mathcal{A}_1^{\dagger}} \otimes \mathcal{H}^{\mathcal{A}_2^{\dagger}}$). 

}, and $\epsilon > 0 $ is sufficiently small such that $M'$ is a valid channel (i.e. such that it is positive). For this pair of channels,  \eqref{e:new condition} becomes
\begin{equation}
T^{B_3}_{B_2}\bullet X^{B_2}_{B_1}\bullet \eta =  I^{B_3}_{B_2}\bullet X^{B_2}_{B_1}\bullet \eta
\end{equation}
However, $X$ is an arbitrary vector, and just as before, the one way that this can hold in all cases is if it holds at the level of the state, i.e.
\begin{equation}
T^{B_3}_{B_2} \bullet \eta =  I^{B_3}_{B_2} \bullet \eta
\end{equation}
That is, the only states that satisfy \eqref{e:b} are those such that the throw-away-and-replace channel applied on the post-selected state of Bob leaves the state invariant. 

A completely equivalent line of reasoning, starting from \eqref{e:d} (i.e. interchanging the role of Alice and Bob), leads directly to the symmetric requirement 
\begin{equation}
T^{A_3}_{A_2} \bullet \eta =  I^{A_3}_{A_2} \bullet \eta.
\end{equation}
Combining these two conditions, we finally arrive at
\begin{equation}
(T^{A_3}_{A_2} \otimes T^{B_3}_{B_2}) \bullet \eta =  (I^{A_3}_{A_2}\otimes I^{B_3}_{B_2}) \bullet \eta
\end{equation}
The state on the left-hand-side has the form $\rho^{A_1 B_1} \otimes \ident_{A_2}\otimes \ident_{B_2}$, while on the right-hand-side we recognise that $(I^{A_3}_{A_2}\otimes I^{B_3}_{B_2}) \bullet \eta_{A_3 B_3}^{A_1 B_1} = \eta_{A_2B_2}^{A_1 B_1}$, and therefore we prove the claim that the only allowed states are those of the form 
\begin{equation} 
\eta^{A_1 B_1}_{A_3 B_3}= \rho^{A_1 B_1} \otimes \ident_{A_3}\otimes \ident_{B_3}.
\end{equation}
\proofend

\noindent This form implies  that
\begin{align}
(L \bullet J_a &\otimes M \bullet K_b)\bullet \eta^{A_1 B_1}_{A_3 B_3} \nonumber \\ &= (L \bullet J_a \otimes M \bullet K_b)\bullet \rho^{A_1 B_1} \otimes \ident_{A_3}\otimes \ident_{B_3} \nonumber \\
&= (J_a \otimes K_b)\bullet \rho^{A_1 B_1} \otimes \ident_{A_2}\otimes \ident_{B_2}
\end{align}
where we used the fact that $L^{A_3}_{A_2}\bullet \ident_{A_3} = \ident_{A_2}$ and $M^{B_3}_{B_2}\bullet \ident_{B_3} = \ident_{B_2}$ for all $L$ and $M$. Thus, the probabilities are independent of the channels $L$ and $M$, and hence
\begin{equation}
p(a,b|x,y) = p(a,b|x',y')
\end{equation}
for all $a$, $b$, $x$, $y$, $x'$, $y'$, which are the sufficient additional conditions satisfied by classical correlations.

Finally, for the case of definite timings but where Alice's experiment is performed before Bob's, we can show the following:

\begin{theorem} \label{thm:constantdenom} \textbf{ A two-party, two-time state obeys the NBTS conditions with a fixed relative time ordering where both of Alice's experiments occur before Bob's if and only if it is proportional to (and thus physically equivalent to) a state where Bob's post-selection is trivial.} If the two-time state $\eta= \mathlarger{\eta}^{A_1 B_1}_{A_3 B_3}$ obeys no-backwards-in-time-signalling in this setting, then
\begin{equation} 
\eta^{A_1 B_1}_{A_3 B_3}= \eta^{A_1 B_1}_{A_3} \otimes  \ident_{B_3}
\,. 
\end{equation} 
\end{theorem} 
\proof 
The NBTS conditions in this case are
\begin{subequations}
\begin{align}
(L \bullet J_a \otimes M )\bullet \eta &= (L' \bullet J_a \otimes M) \bullet 
\eta \label{e:a1}\\
(L \bullet J_a \otimes M )\bullet \eta &= (L \bullet J_a \otimes M') \bullet 
\eta \label{e:b1}\\
%\\
(L \otimes M \bullet K_b )\bullet \eta &= (L \otimes M' \bullet K_b) \bullet \eta \label{e:c1}
%(L \otimes M \bullet K_b ) \bullet \eta &= (L' \otimes M \bullet K_b) \bullet \eta \label{e:d}
\end{align} 
\end{subequations}
where $J_a$ is Alice's measurement; $K_b$  Bob's measurement, and $ L, L', M, M'$ are completely positive trace preserving maps. 
The proof follows immediately from the previous proof. In particular, by the same logic as previously it is still the case that 
\begin{equation}\label{eq:Alicelocal}
T^{B_3}_{B_2} \bullet \eta =  I^{B_3}_{B_2} \bullet \eta
\end{equation}
from which the claim follows. 
Furthermore, it must be the case that $\eta^{A_1 B_1}_{A_3} $ is constrained such that the full state $\eta^{A_1 B_1}_{A_3 B_3}$ is a linear two-time state. As shown in \cite{Silva2017}, the four necessary and sufficient conditions for a two-time state to be a process matrix are also necessary and sufficient conditions for a two-time state to be linear. These are:
\begin{subequations}\label{e:ew conditions}
\begin{align}
	\left( \mathlarger{\mathbb{I}}_{\mathsmaller{\oA_1}}\!
	\otimes
		\mathlarger{\mathbb{I}}^{\mathsmaller{\oA_3}}\!
			\otimes
				\mathlarger{\mathbb{I}}_{\mathsmaller{\oB_1}}\!
					\otimes
						\mathlarger{\mathbb{I}}^{\mathsmaller{\oB_3}}\! \right)
							\bullet
								\eta
									&= d_{\oA_3}d_{\oB_3}, \label{eq:lin1}\\
	(\chan{I}{\oA_2}{\oA_3} 
		\otimes \chan{T}{\oB_1}{\oB_3})
			 \bullet 
			 	\eta &= 
					(\chan{T}{\oA_2}{\oA_3} 
						\otimes
					 \chan{T}{\oB_1}{\oB_3} )
					 	\bullet 
					 		\eta, \label{eq:lin2}\\
	(\chan{T}{\oA_1}{\oA_3} 
		\otimes
			\chan{I}{\oB_2}{\oB_3} )
			\bullet 
				\eta
					&=( \chan{T}{\oA_1}{\oA_3} 
						\otimes
						\chan{T}{\oB_2}{\oB_3} )
							\bullet
								 \eta, \label{eq:lin3}\\
	(\chan{I}{\oA_2}{\oA_3} 
		\otimes
			\chan{I}{\oB_2}{\oB_3})
				 \bullet 
				 	\eta &= 
				 		(\chan{I}{\oA_2}{\oA_3}
				 			 \otimes
				 			 	\chan{T}{\oB_2}{\oB_3})\bullet 
				 			 		\eta  \nonumber
				 			 		\\+ 
				 			 		( \chan{T}{\oA_2}{\oA_3}
				 			 			\otimes
&				 			 			\chan{I}{\oB_2}{\oB_3})\bullet \eta
		  - (\chan{T}{\oA_2}{\oA_3} 
			\otimes
				\chan{T}{\oB_2}{\oB_3})
					\bullet 	
					\eta
					\label{eq:lin4}
\end{align}
\end{subequations}
of which Eqns.~\eqref{eq:lin1}, \eqref{eq:lin3} and \eqref{eq:lin4} are trivially satisfied by a state of the form $\eta^{A_1 B_1}_{A_3} \otimes  \ident_{B_3}$ (using $T^{B_3}_{B_X}\bullet \ident_{B_3} = \ident_{B_X}$). Eq.~\eqref{eq:lin2} is satisfied with the additional constraint that Alice cannot locally signal backwards in time, i.e. Eq.~\eqref{eq:Alicelocal}. Together this implies that $\eta^{A_1 B_1}_{A_3 B_3}$ is linear. \proofend

Once again, from this form it directly follows that
\begin{align}
(L \bullet J_a &\otimes M \bullet K_b)\bullet \eta^{A_1 B_1}_{A_3 B_3} \nonumber \\ &= (L \bullet J_a \otimes M \bullet K_b)\bullet \eta^{A_1 B_1}_{A_3}\otimes \ident_{B_3} \nonumber \\
&= (L \bullet J_a \otimes K_b)\bullet\eta^{A_1 B_1}_{A_3}\otimes \ident_{B_2} 
\end{align}
which is independent of $M$. Therefore, we have
\begin{equation}
p(a,b|x,y) = p(a,b|x,y')
\end{equation}
for all $a$, $b$, $x$, $y$, $y'$. Again, these are sufficient conditions (on top of NBTS) that guarantee that a correlation is classical. Hence the set of two-time states cannot generate non-classical correlations. 

\end{document}